\documentclass[aip,twocolumn, showpacs,superscriptaddress,graphicx]{revtex4}
\usepackage{amssymb}
\usepackage{graphicx}
\usepackage{amsmath}
\usepackage{amsfonts}
\usepackage{epsfig}
\usepackage{epstopdf}

\graphicspath{{ffunc/}{figures/}}
\draft

\begin{document}

\title{Observability of surface currents in p-wave superconductors}
\author{S.~V.~Bakurskiy}
\affiliation{Skobeltsyn Institute of Nuclear Physics, Lomonosov Moscow State University,
Moscow 119991, Russian Federation}
\affiliation{Moscow Institute of Physics and Technology, Dolgoprudny, Moscow Region,
141700, Russian Federation}
\author{N.~V.~Klenov}
\affiliation{Moscow Institute of Physics and Technology, Dolgoprudny, Moscow Region,
141700, Russian Federation}
\affiliation{Faculty of Physics, Lomonosov Moscow State University, 
Moscow 119992, Russian Federation}
\affiliation{All-Russian Research Institute of Automatics n.a. N.L. Dukhov (VNIIA), 127055, Moscow, Russia}
\author{I.~I.~Soloviev}
\affiliation{Skobeltsyn Institute of Nuclear Physics, Lomonosov Moscow State University,
 Moscow 119991, Russian Federation}
\affiliation{Moscow Institute of Physics and Technology, Dolgoprudny, Moscow Region,
141700, Russian Federation}
\author{M.~Yu.~Kupriyanov}
\affiliation{Skobeltsyn Institute of Nuclear Physics, Lomonosov Moscow State University,
Moscow 119991, Russian Federation}
\affiliation{Moscow Institute of Physics and Technology, Dolgoprudny, Moscow Region,
141700, Russian Federation}
\author{A.~A.~Golubov}
\affiliation{Moscow Institute of Physics and Technology, Dolgoprudny, Moscow Region,
141700, Russian Federation}
\affiliation{Faculty of Science and Technology and MESA+ Institute for Nanotechnology,
University of Twente, 7500 AE Enschede, The Netherlands}
\date{\today }
\date{\today }

\begin{abstract}

A general approach is formulated to describe spontaneous surface current distribution in a chiral p-wave superconductor. We use the quasiclassical Eilenberger formalism in the Ricatti parametrization to describe various types of the superconductor surface, including arbitrary roughness and metallic behaviour of the surface layer. We calculate angle resolved distributions of the spontaneous surface currents and formulate the conditions of their observability. We argue that local measurements of these currents by muSR technique may provide an information on the underlying pairing symmetry in the bulk superconductor.

\end{abstract}

\maketitle

\section{Introduction\label{Intro}}

There is currently growing interest in theoretical and experimental studies
of topologial superconductors \cite{Kallin}-\cite{MS}. One of the well-known
examples is $Sr_2RuO_4$, which is thought to be a chiral triplet p-wave
superconductor \cite{Maeno}-\cite{Maeno2016}. There are several features of 
triplet p-wave superconductivity, among them the key signature is appearence of the odd-frequency states
in the vicinity of surface, leading to formation undergap bound states \cite{Zhang}-\cite{LuBo}. The chiral
superconductivity is highlighted by existence of the spontaneous
surface currents being one of most interesting among the predicted features. Unfortunately,
these currents have never been experimentally detected yet \cite{PGB},
\cite{Kirtley}.

One of possible reasons for the absence of the surface currents could be the
presence of non-superconducting layers on the surface of the samples, which
may be formed due to sample degradation. Such layers can be characterized by
various degrees of diffusivity and by range of thickness, $d,$ measured in
superconducting decay length units, $\xi _{0}$. 
It was demonstrated recently within the quasiclassical Eilenberger formalism
\cite{Eilenberger} that the edge currents in a chiral p -wave
superconductor are robust with respect to surface roughness \cite{Suzuki} in the regime of small thickness $d$. On the other hand, calculations done
for the case of specular surface within the tight-binding model \cite{Ashby}-%
\cite{Huang} have shown that the spontaneous currents could be suppressed if $d$ is of the same order or greater than the decay length $\xi _{0}$.

In order to establish general conditions for the existence of surface currents in chiral p-wave superconductors, in this work we formulate
the general approach in the framework of the Eilenberger formalism \cite%
{Eilenberger} for various experimental realizations of surface layers in
these materials. We investigate thin $(d\ll \xi _{0})$ and
thick $(d\gg \xi _{0})$ clean and diffusive films in proximity with bulk
p-wave chiral superconductor. We show that, in accordance with the results of
\cite{Suzuki}, spontaneous currents are not sensitive to surface roughness
in the limit of thin surface layer, while in the opposite case $(d\gg \xi
_{0})$ these currents are strongly suppressed. We also discuss the peculiarities of
current distributions inside the sample for different types of surface layers and the possibility of their experimental observation.

\section{Model\label{Sec1}}

The description of the surface properties of chiral p-wave superconductor
has been done within the framework of the quasiclassical Eilenberger equations
\cite{Eilenberger}. To solve the problem, we will assume that the conditions
of the clean limit (scattering time $\tau \rightarrow \infty $ ) are valid
in the bulk superconductor region $-\infty<x\leqslant0$ and the equations
have the form
\begin{equation}
2\omega f(x,\theta )+v\cos (\theta )\frac{d}{dx}f(x,\theta )=2\Delta
g(x,\theta ),  \label{El0}
\end{equation}%
\begin{equation}
2\omega f^{+}(x,\theta )-v\cos (\theta )\frac{d}{dx}f^{+}(x,\theta )=2\Delta
^{\ast }g(x,\theta ),  \label{El0a}
\end{equation}%
\begin{equation}
2v\cos (\theta )\frac{d}{dx}g_{\omega }(x,\theta )=2\left( \Delta ^{\ast
}f_{\omega }-\Delta f_{\omega }^{+}\right) .  \label{El0b}
\end{equation}%
Here $g(x,\theta )$, $f(x,\theta )$ and $\ f^{+}(x,\theta )$ are normal and
anomalous Eilenberger functions, $\theta $ is angle between the vector
normal to the interface and the direction of the electron Fermi velocity $v;$
$\omega =\pi T(2n+1)$ are Matsubara frequencies and $T$ is temperature, $x$
is coordinate along the axis normal to the boundary. Pair potential $\Delta
(x,\theta )$ in a chiral p-wave superconductor is separated into two
terms $\Delta =\Delta _{x}\cos (\theta )+i~\Delta _{y}\sin (\theta )$.

Equations (\ref{El0})-(\ref{El0b}) should be supplemented with
the self-consistency equation, which can be written in the form \cite{Bruder}
\begin{eqnarray}
\Delta _{x}\ln \frac{T}{T_{c}}+2\pi T\sum_{\omega }\frac{\Delta _{x}}{\omega
}-\left\langle 2\cos (\theta ^{\prime }) Re f(x,\theta ^{\prime
})\right\rangle &=&0,  \label{Sc0b} \\
\Delta _{y}\ln \frac{T}{T_{c}}+2\pi T\sum_{\omega }\frac{\Delta _{y}}{\omega
}-\left\langle 2\sin (\theta ^{\prime }) Im f(x,\theta ^{\prime
})\right\rangle &=&0.  \label{Sc0c}
\end{eqnarray}
Here $\left\langle ...\right\rangle =(1/2\pi )\int_{0}^{2\pi }(...)d\theta $
and $T_{c}$ is the critical temperature.

Previously, the properties of the superconductor surface have been described \cite%
{Golubov1}-\cite{Golubov2} in the frame of Ovchinnikov model \cite%
{Ovchinnikov}. Within this model it is assumed that the surface is covered by a thin
diffusive normal metal layer. In the present study we go beyond this approximation
and consider the situation when a clean p-wave superconductor is covered
by a normal metal of finite thickness
with respect to the coherence length $\xi _{0}=v/2\pi T_{c} $ and electron mean free
path $l_{e}$ and having mirror-reflecting surface. Inside this layer, located in the area $0\leq x\leq d$,
Eilenberger equations \cite{Eilenberger} transform to

\begin{equation}
v\cos (\theta )\frac{d}{dx}f(x,\theta )=\frac{1}{\tau }\left( g\left\langle
f\right\rangle -f\left\langle g\right\rangle \right) -2\omega f(x,\theta ),
\label{El2a}
\end{equation}%
\begin{equation}
v\cos (\theta )\frac{d}{dx}f^{+}(x,\theta )=\frac{1}{\tau }\left(
g\left\langle f^{+}\right\rangle -f^{+}\left\langle g\right\rangle \right)
+2\omega f^{+}(x,\theta ),  \label{El2b}
\end{equation}%
\begin{equation}
2v\cos (\theta )\frac{d}{dx}g(x,\theta )=\frac{1}{\tau }\left( f\left\langle
f^{+}\right\rangle -f^{+}\left\langle f\right\rangle \right).  \label{El2c}
\end{equation}%
Following the procedure developed in \cite{Golubov1}-\cite{Golubov2} we will
consider (\ref{El2a})-(\ref{El2c}) as the system of linear equations in
which $\left\langle f\right\rangle $, $\left\langle f^{+}\right\rangle $ and
$\left\langle g\right\rangle $ should be determined self-consistently in the
process of finding a solution to the equations (\ref{El0})-(\ref{El2c}). To
simplify the problem further we introduce pairing functions of the incident $%
f_{+}(\theta )=f(\theta )$ and reflected $f_{-}(\theta )=f(\pi +\theta )$
particles, which are defined in the segment of angles $-\pi /2\leq \theta
\leq \pi /2.$ 
The functions $\ f_{\pm }^{+}$ and $\ g_{\pm }$ are defined below in a
similar manner.

For the development of numerical algorithms for solution of the Eilenberger
equations it is convenient to rewrite them using the Ricatti parametrization
\cite{Schopohl,Tanaka6}.
\begin{equation}
f_{\pm }=\frac{2a_{\pm }}{1+a_{\pm }b_{\pm }},\ f_{\pm }^{+}=\frac{2b_{\pm }%
}{1+a_{\pm }b_{\pm }},\ g_{\pm }=\frac{1-a_{\pm }b_{\pm }}{1+a_{\pm }b_{\pm }%
}.  \label{Ric1}
\end{equation}%
For the Ricatti functions $a_{\pm }$ and $b_{\pm }$ from (\ref{El0})-(\ref%
{El2c}) we have
\begin{eqnarray}
v\cos (\theta )\frac{d}{dx}a_{\pm } &=&\Delta -\Delta ^{\ast }a_{\pm
}^{2}\mp 2\omega a_{\pm },  \label{El_ric_cl_A} \\
v\cos (\theta )\frac{d}{dx}b_{\pm } &=&-\Delta ^{\ast }+\Delta b_{\pm
}^{2}\pm 2\omega b_{\pm },  \label{El_ric_cl_B}
\end{eqnarray}%
in the clean superconducting region and
\begin{eqnarray}
\frac{da_{\pm }}{dx} &=&\mp \frac{\left\langle f\right\rangle -a_{\pm
}^{2}\left\langle f^{+}\right\rangle -2a_{\pm }\left\langle g\right\rangle }{%
2\tau v\cos (\theta )}\mp \frac{2\omega a_{\pm }}{v\cos (\theta )},
\label{El_ric_dr_A} \\
\frac{db_{\pm }}{dx} &=&\mp \frac{\left\langle f^{+}\right\rangle -b_{\pm
}^{2}\left\langle f\right\rangle -2b_{\pm }\left\langle g\right\rangle }{%
2\tau v\cos (\theta )}\pm \frac{2\omega b_{\pm }}{v\cos (\theta )},
\label{El_ric_dr_B}
\end{eqnarray}%
in the diffusive layer. The subscript $+$ in (\ref{Ric1})-(\ref{El_ric_dr_B}%
) indicates that functions are calculated along trajectories in the
direction towards the SN boundary, while the subscript $-$ denotes the
functions calculated on trajectories in the direction away from the SN interface
towards the bulk superconductor.  Far from the SN interface (at $x\rightarrow
-\infty $) the values of $a_{\pm }$ and $b_{\pm }$ equal to their magnitudes
in a homogeneous superconductor
\begin{equation}
a_{\pm }=\pm \frac{\Delta }{\omega +\sqrt{\omega ^{2}+|\Delta |^{2}}},
\label{As0}
\end{equation}%
\begin{equation}
b_{\pm }=\pm \frac{\Delta ^{\ast }}{\omega +\sqrt{\omega ^{2}+|\Delta |^{2}}}%
,  \label{As1}
\end{equation}%
where $\Delta $ is the bulk value of the pair potential.

Finally, the problem must be supplemented by the boundary conditions \cite%
{Bakurskiy}  at the free surface of the diffusion layer $(x=d)$%
\begin{equation}
b_{+}(d,-\theta )=b_{-}(d,\theta ),  \label{BC1a}
\end{equation}%
\begin{equation}
a_{-}(d,-\theta )=a_{+}(d,\theta ),  \label{BC1b}
\end{equation}
and by the requirement of continuity of the $a_{\pm }$ and $b_{\pm }$ functions
on the SN interface. 

The boundary problem (\ref{El_ric_cl_A})-(\ref%
{BC1b}) has been solved numerically.
To solve it, we start with some trial functions for order parameter, $%
\left\langle f\right\rangle,$ $\left\langle f^{+}\right\rangle,$ $%
\left\langle g\right\rangle$ and intergrade numerically equations (\ref%
{El_ric_cl_A}) - (\ref{El_ric_dr_B}) for $a_{+}(x,\theta )$ and $%
b_{-}(x,-\theta )$ moving along the trajectory from the bulk values of these
functions (\ref{As0}) at infinity $(x=-\infty )$ towards the free N layer
surface $(x=d)$ making use of the requirement of continuity of the $a_{+}$
and $b_{-}$ functions on the SN interface $(x=0)$. Thus obtained solutions $%
a_{+}(0,\theta )$ and $b_{-}(0,-\theta )$ and boundary conditions (\ref{BC1a}%
), (\ref{BC1b}) determine the magnitudes of $a_{-}(0,\theta )$ and $%
b_{+}(0,\theta )$. Starting from these solutions, the functions $a_{-}(x,\theta )$ and $%
b_{+}(x,\theta )$ are obtained by integration along the trajectories going
across the diffusive N layer into the bulk superconductor.  The averaged
functions $\left\langle f\right\rangle,$ $\left\langle f^{+}\right\rangle,$ $%
\left\langle g\right\rangle$ and the spatial dependence of the order
parameter $\Delta (x)$ are determined in an iterative self-consistent way
using the definitions (\ref{Ric1}) and self-consistency equations ( (\ref%
{Sc0b})-(\ref{Sc0c}).

The calculated Ricatti functions $a_{\pm }(x,\theta )$ and $b_{\pm }(x,\theta)
$ permit one to restore the angular distribution of the Eilenberger functions in
any cross-sectional plane inside SN bilayer and calculate the spatial
distribution of the density of spontaneous edge current $j(x)$
\begin{equation}
j(x)=2evN_{0}T\sum_{\omega >0}j_{\omega}(x), \
j_{\omega}(x)=\int\limits_{0}^{2\pi }j_{\omega}(x, \theta)\sin \theta
d\theta,  \label{current}
\end{equation}
where $N_{0}$ is the density of states and angular resolved supercurrent
density $j_{\omega}(x, \theta)=Im(g_{\omega }(x, \theta))=Im(\sqrt{%
1-f_{\omega }(\theta )f_{\omega }(\pi -\theta )})$. When writing the last
equality we used
the normalization condition for triplet functions
\begin{equation}
g_{\omega }^{2}(\theta )-f_{\omega }(\theta )f_{\omega }^{+}(\theta )=1,
\label{norcond}
\end{equation}
and the symmetry relation%
\begin{equation}
f_{\omega }(\theta )=-f_{\omega }^{+}(\pi -\theta ).  \label{symrel}
\end{equation}%
For simplicity we will suppose further that the amplitudes $f_{\omega }(x,
\theta )$ are small, and making use of (\ref{norcond}), (\ref{symrel})
we rewrite (\ref{current}) in the form
\begin{equation}
\begin{array}{c}
j_{\omega }(x)=\int\limits_{-\pi/2}^{\pi/2 }\left( Re(f_{\omega }(x,\theta
))Im(f_{\omega }(x,\pi -\theta ))+\right. \\
\left. +Re(f_{\omega }(x,\pi -\theta ))Im(f_{\omega }(x,\theta ))\right)
\sin \theta d\theta ,%
\end{array}
\label{jsmall}
\end{equation}

It is seen from Eq.(\ref{jsmall}) that  the appearance of the spontaneous
surface current is a result of superposition of anomalous Eilenberger
functions of incident $f_{\omega }(x,\theta )$ and reflected $f_{\omega
}(x,\pi -\theta )$ from the free interface trajectories. In the geometry considered in this work, when the $p_{x}$ principal axis is parallel to the interface normal, there is  general relation $f_{\omega }(\theta )=f_{\omega }^{\ast }(-\theta
)$ providing  the odd symmetry of imaginary part of pair amplitude. Far from the
SN interface inside the bulk p-wave superconductor there is the odd
symmetry of real part of pairing function $Re(f_{\omega }(\theta ))=$ $%
-Re(f_{\omega }(\pi -\theta ))$ and the terms under integration in (\ref%
{jsmall}) have the same magnitude, but different sign and cancel each other
during integration resulting in the absence of a supercurrent. However, in
the vicinity of the surface of p-wave superconductor or SN interface the
symmetry of $Re(f_{\omega }(\theta ))$ on ingoing and outgoing trajectories is broken resulting in generation of a spontaneous supercurrent.

\section{Spontaneous current \label{Current}}

In this Section we present the results of calculations of spatial distributions of the pair potential $\Delta (x),$, the spontaneous supercurrent $j(x),$ as well as the angular resolved pair
amplitude $f_{1}(x,\theta)$ and the density of the first spectral component $%
j_{1}(x)$ in (\ref{current})  together with its angular resolved
distribution $j_{1}(x,\theta)$ at $\omega=\pi T$ for four different
situations.
They are specular p-wave surface ($d\ll \ell_{e}$, $d\ll \xi_{0}$); rough
surface ($d\gg \ell_{e}$, $d\ll \xi_{0}$); clean metallic surface ($d\geq
\xi_{0}$, $\ell_{e}\gg \xi_{0}$) and diffusive metallic surface ($d\geq
\xi_{0}$, $\ell_{e}\ll \xi_{0}$).

All  calculations have been performed at temperature $T=0.5T_{C}$ and the
results are shown in Fig.\ref{F_Specular}-Fig.\ref{f_inter}. Numerical analysis shows that at $T=0.5T_{C}$ the calculated value of $j_{1}(x)$ provides the contribution to the full current with accuracy of the order of 10 percent, therefore all the features discussed below reflect the behavior of total spontaneous current.

The figures
consist of several panels. Panel (a) shows the coordinate dependencies of the pair
potential for a chiral $p_{x}+ip_{y}$ superconductor. Panels (b) and (c)
demonstrate the density of the first spectral component $j_{1}(x)$ in (\ref%
{current})  and its angular resolved distribution $j_{1}(x,\theta)$,
respectively. Panels (d) show real (solid lines) and imaginary (dashed lines)
parts of $f_{1}(x,\theta)$ calculated at different distances $x/\xi_0$ from
the SN interface. Color in panels d) specifies the information on the sign
of $f_{1}(x,\theta)$ functions: the red color corresponds to the positive
values of $f_{1}(x,\theta)$, while the blue one represents their negative
magnitudes.

\bigskip \emph{Specular p-wave surface ($d\ll \ell_{e}$, $d\ll \xi_{0}$)}.

\bigskip Figure \ref{F_Specular} summarize the results obtained for
specular p-wave surface under absence of any clean or diffusive N layer on
the surface.

In full agreement with the previously obtained results \cite{Bakurskiy} in
this case bulk pair potential (see Fig\ref{F_Specular}a) has BCS amplitude
(at our temperature $\Delta _{x}=\Delta _{y}\approx 1.67T_{C}$) due to
spherical symmetry. The $\Delta _{y}$ component increases as we approach the
surface and grows up to bulk value for $p_{y}$ symmetry since after
reflection electrons propagate into the band with the same sign of pair
potential. The corresponding part of angular resolved distribution of pair
amplitude $f_{1}(x,\theta)$ (see dashed lines in Fig. \ref{F_Specular}d)
demonstrate a similar behavior. The magnitudes of this imaginary parts of $%
f_{1}(x,\theta)$ only slightly increase with $x$ decrease in the full
accordance with $\Delta _{y}$ growth.

Contrary to that, in the vicinity of interface $\Delta _{x}$ is suppressed up
to zero due to reflection of electrons from the band having positive sign of
pair potential into the band with negative one. The transformations of real
part of angular resolved distribution of pair amplitude $f_{1}(x,\theta)$
(see solid lines in Fig. \ref{F_Specular}d) is more complex. Starting from
the bulk (red curve in Fig. \ref{F_Specular}d(0)) we first exhibit the
suppression of this component with $x$ decrease, as shown in Fig. \ref%
{F_Specular}d(1)) - Fig. \ref{F_Specular}d(3)). This is a direct consequence
of the suppression of the $\Delta _{x}$ component of the order parameter. At
the surface the boundary conditions (\ref{BC1a}), (\ref{BC1b}) dictate for
the outgoing functions $f_{1}(0,\theta)$ a sign opposite to that existing in
the depth of the superconductor. This perturbation relaxes on characteristic
scale $\xi=\xi_{0} \cos(\theta)(\pi T_{C}/ \sqrt{\omega^{2}+|\Delta |^{2}} )$
of the equation (\ref{El_ric_cl_A}), (\ref{El_ric_cl_B}), which is $\theta$
dependent. It is for this reason the sign of $f_{1}(0,\theta)$ functions
recovers the faster the smaller is $\cos(\theta)$, that is the larger is
deviation of outgoing trajectory from the normal to the superconductor
surface direction. This is clearly seen in Fig. \ref{F_Specular}d(3). It
demonstrates that at $x=-0.2\xi_0$ the sign of functions $f_{1}(-0.2
\xi_0,\theta)$ is different for different $\theta$. At $x=-0.4\xi_0$ the
pairing functions are negative in all angle domain. Further decrease of $x$
results in increase of magnitude of $f_{1}(x,\theta)$ and to the full
relaxation to the balk values at $x=-2\xi_0.$

This lack of symmetry on the trajectories incoming and outgoing from the surface
leads to generation of spontaneous supercurrent. The angular
resolved density of its first spectral component $j_{1}(x,\theta)$ is shown
in Fig. \ref{F_Specular}c. In accordance with (\ref{current}) its
integration over $\theta$ results in nonzero $j_{1}(x),$ which attains a
maximum at the surface and monotonically decays into superconductor.
Exactly at the surface plane ($x=0$) the incomig and reflecting electrons make
equal contributions to $j_{1}(0).$ The component of $j_{1}(x)$ from the
outgoing electrons decreases monotonically to zero with the decrease of $x$ and
becomes negative in the bulk area resulting in full compensation of the
positive part of $j_{1}(x)$ generated by incoming particles. The maximum of $%
j_{1}(x,\theta)$ dependencies is achieved at $\theta \approx \pi/4$
resulting in decay scale of $j_{1}(x)$ of the order of $\xi_0 / \sqrt{2}.$

\bigskip \emph{Rough surface ($d\gg \ell_e$ and $d\ll \xi_0$.})

\bigskip  The situation described above changes dramatically when there is
strong diffuse scattering of electrons at the free surface of the
superconductor (see Fig. \ref{f_rough}). Below we model the diffuse
scattering of electrons by a normal layer placed on the superconductor
surface. This layer has thickness $d$ equal to electronic mean free path $%
\ell_e,$ decay length $\xi_0=3 \ell_e$ and $\Delta=0$.

Figure \ref{f_rough}a shows that in the considered situation in the vicinity
of SN interface there is suppression of both components of the order
parameter. The panels Fig.\ref{f_rough}d(0) - Fig.\ref{f_rough}d(5)
demonstrate that amplitudes of angular resolved distributions of real (solid
red curves) and imaginary (dashed curves) parts of functions $f_{1}(x,\theta)
$ for ingoing trajectories ($-\pi/2 \leq \theta \leq \pi/2)$ decrease slowly
then $x$ goes to SN interface located at $x=0$. In the diffusive layer, the
imaginary part of $f_{1}(x,\theta)$ decays rapidly due to averaging of
its alternating parts.  At $x=d$ (see Fig.\ref{f_rough}d(8)) the imaginary
part of $f_{1}(d,\theta)$ practically disappear and boundary conditions (\ref%
{BC1a}), (\ref{BC1b}) provide nearly isotropic angular distribution of $%
f_{1}(d,\theta)$. The propogation on outlet trajectories from free surface
back to SN interface across the diffusive layer results in further
isotropisation of $f_{1}(x,\theta)$, as it follows from Fig.\ref{f_rough}%
d(5).

During the propagation into the S film (in the angular domain $\pi/2 \leq x
\leq 3\pi/2$) the isotropic function $f_{1}(0,\theta)$ distribution rapidly vanished
due to strong nucleation of the imaginary component of $f_{1}(x,\theta)$. As
it follows from Fig.\ref{f_rough}d(4) - Fig.\ref{f_rough}d(2) it recovers
its bulk distribution practically at $x=-0.5\xi_0$. Evolution of real part
of $f_{1}(x,\theta)$ is nearly the same as discussed in the previous
paragraph. First it goes to zero with $x$ decrease, then it changes its sign and
tends monotonically to its bulk behavior.

Figures \ref{f_rough}b and Fig. \ref{f_rough}c show the spatial variation of the
first spectral component $j_{1}(x)$ and its angular resolved spectral
component $j_{1}(x,\theta)$ in the considered regime. It is seen that
spontaneous supercurrent is pushed out from the N layer towards the SN
interface. It reaches its maximum at the SN boundary and then decreases with
the distance deep into the superconductor. Note, that there are two
characteristic scales in this decay. In order to understand these phenomena
it is enough to look at Fig. \ref{f_rough}c and use the expression (\ref%
{jsmall}). Expression (\ref{jsmall}) reads that in the N layer and in a
vicinity of SN interface the main contribution to $j_{1}(x)$ provided by the
product of isotropic real part of of $f_{1}$ ($p_{x}$ component) on outgoing
trajectories and imaginary part of $f_{1}$ ($p_{y}$ component) on ingoing
trajectories. The latter is strongly suppressed in the N film (see Fig. \ref%
{f_rough}d(6) - Fig. \ref{f_rough}d(8)) resulting in the current
suppression. At the SN interface, the $p_{y}$ component of $f_{1}$ on ingoing
trajectories still exists. It achieves maximum values at $\theta \approx
70^{o}$ thus providing the maximum in the angular resolved spectral
component $j_{1}(x,\theta)$ at $\theta \approx 75^{o}$. Thus, in the
vicinity of the SN boundary main contribution to the spontaneous current
comes from trajectories nearly parallel to interface. The characteristic
decay length of the anomalous functions on these trajectories is $%
\xi_{0}\cos(75^{o})$, that is much smaller than $\xi_{0}.$ Further deviation
from the SN boundary is accompanied by deformations in the $f_{1}(x,\theta)$,
which will eventually lead to recovery of $f_{1}(x,\theta)$ to the bulk one.
These modifications are accompanied by reconstruction of $j_{1}(x,\theta)$ and by displacement of the
position of its maximum to angles close to $\pi/4$. As a result, for
smaller $x$ the decay length of $j_{1}(x)$  goes to the same value $%
\xi_{0}\cos(\pi/4)$ that has been found for the specular superconductor
interface.

In the case where the thickness of the N layer becomes comparable with the decay length $\xi_0$ there is
significant attenuation of superconducting correlations in this layer
resulting in strong suppression of superconductivity at outgoing from SN
interface trajectories. This, in turn, leads to further displacement of
spontaneous supercurrents into the bulk region. We will demonstrate this
effect below by considering two examples; "clean metallic surface"
and "diffusive metallic surface". In both cases we will assume that there is
no intrinsic superconductivity in the N metal and that its thickness and electronic
mean free path are $d=\xi_0$, $l_e=\infty$ in the clean case and $d= l_e
$ and $\xi_0=0.3 l_e$ in the diffusive case.

\bigskip

\emph{Clean metallic surface ($d\geq \xi_{0}$, $\ell_{e}\gg \xi_{0}$).}

\bigskip

Figure \ref{f_clean} summarizes the results obtained for the clean metallic
surface. The panels Fig.\ref{f_clean}d(0) - Fig.\ref{f_clean}d(5)
demonstrate that amplitudes of angular resolved distributions of real (solid
red curves) and imaginary (dashed curves) parts of functions $f_{1}(x,\theta)
$ for inlet trajectories ($-\pi/2 \leq \theta \leq \pi/2)$ behaves nearly in
the same manner as in the cases considered previously. They decrease slowly
then $x$ goes to SN interface located at $x=0$. Propagation into the N layer
leads to full suppression of anomalous functions and they are zero in outlet
trajectories at SN interface (see Fig.\ref{f_clean}d(5)). It is for this
reason the spontaneous supercurrent is zero in the area $0\leq x \leq d$.
From Fig.\ref{f_clean}d(0) - Fig.\ref{f_clean}d(4.5) it is seen that
imaginary part of anomalous functions recover faster compare to real one.
This imbalance provides nucleation of spontaneous currents. Their density
increases (see Fig.\ref{f_clean}b) and achieves its maximum at $x\approx-0.5 \xi_0$%
. At this point (see Fig.\ref{f_clean}d(2)) the imaginary part of anomalous
functions saturates at the bulk value. With further decrease of $x$ the magnitude of the real part of $f_1$ increases and defines the
supercurrent component, which compensates the one from inlet electrons. As a
result, the density of spontaneous current starts to decrease for $x< -0.5
\xi_0$ and goes to zero in the bulk. Figure \ref{f_clean}c shows that
maximum in spectral component $j_{1}(x,\theta)$ is achieved at $\theta\approx
\pi/4$. It means that characteristic scale of $j_{1}(x)$ variation is $\xi_0
\cos(\pi/4))$, similar to the case of specular p-wave
superconductor interface.

\bigskip  \emph{Diffusive metallic surface ($d\geq \xi_{0}$, $\ell_{e}\ll
\xi_{0}$).}

\bigskip

The difference between the diffusive metallic interface from the clean one
lies in the fact that due to the presence of electron scattering centers
near the border, some of electrons can be reflected back at small distances
from the SN interface. This means that in contrast to the clean limit
considered previously, there is a probability for nucleation of small isotropic
diffuse component in $f_1 (0, \theta)$ distribution. Its presence would lead
to the appearance of a diffuse current peak at $x=0$ (see Fig. \ref{f_inter}%
b). The angular resolved density of spontaneous supercurrent $j_{1}(x,\theta)
$ (see Fig. \ref{f_inter}c) clearly demonstrates the existence of two
characteristic angles. It has one maximum at $\theta \approx 75^{o}$
originated by diffusive processes and the second one at $\theta \approx \pi/4$.
This leads to the existence of two characteristic scales in $j_1 (x)$
dependence. They are the "diffusion scale" $\xi_{d}=\xi_0 \cos(75^{o})$ and
"clean scale" $\xi_{c}=\xi_0 \cos(\pi/4).$ It is seen from Fig. \ref{f_inter}%
b) that associated with this scales maximums in $j_1 (x)$ dependence can be
resolved due to large difference between $\xi_{d}$ and $\xi_{c}$ and the
shift of order of $0.5 \xi_{0}$ between the space positions of this
peculiarities.

\section{Discussion \label{Discussion}}

Our studies provide the detailed analysis of the mechanisms leading to the formation of spontaneous currents at the surface of a chiral p-wave superconductor in terms of the Ricatti functions.
 From the structure of the expressions (\ref{current})-(\ref{jsmall}), which determine the magnitude and direction of spontaneous currents it follows that these currents  depend on the products of the real and imaginary parts of the anomalous Green's functions on the incoming and outgoing trajectories from the border. These parts strongly depends
 on the requirements of the symmetry of the Green's functions, which, in turn, dictate the form of the boundary conditions on the open surface. For d$_{x^2-y^2}$-wave superconductors, $b_{\pm }(x,\theta )=a_{\mp }(x,\theta ),$ holds and  the conditions (\ref{BC1a}), (\ref{BC1b}) are reduced to the
relation $b_{+}(d,-\theta )=a_{+}(d,\theta ),$ which under the circumstances that lead to  suppression of the order parameter down to zero at the free interface
  simultaneously ensures the vanishing of anomalous functions on that surface   \cite{Golubov2}. 

\begin{figure*}[h]
\begin{minipage}[h]{0.5\linewidth}
\begin{minipage}[h]{0.99\linewidth}
\center{\includegraphics[width=1\linewidth]{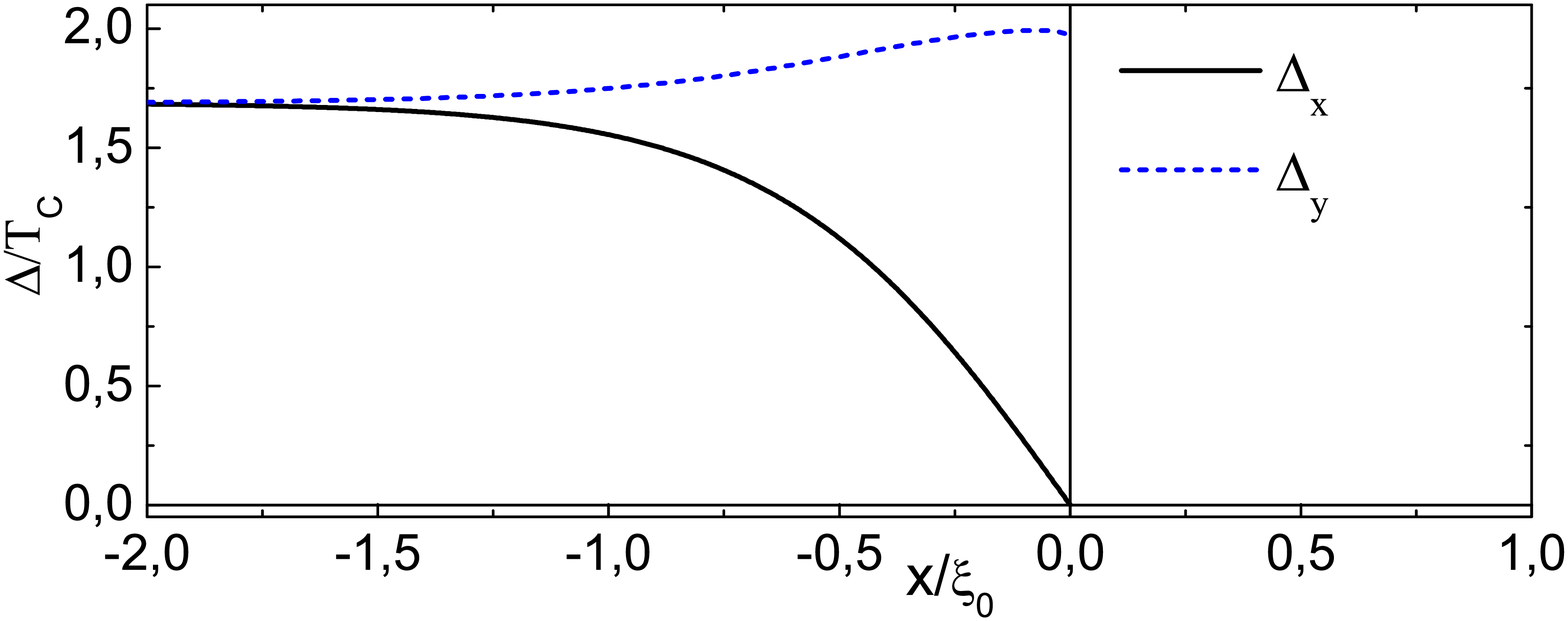}} \\
\vspace{-2 mm}
\raggedright{a)}
\end{minipage}
\vfill
\begin{minipage}[h]{0.99\linewidth}
\center{\includegraphics[width=1\linewidth]{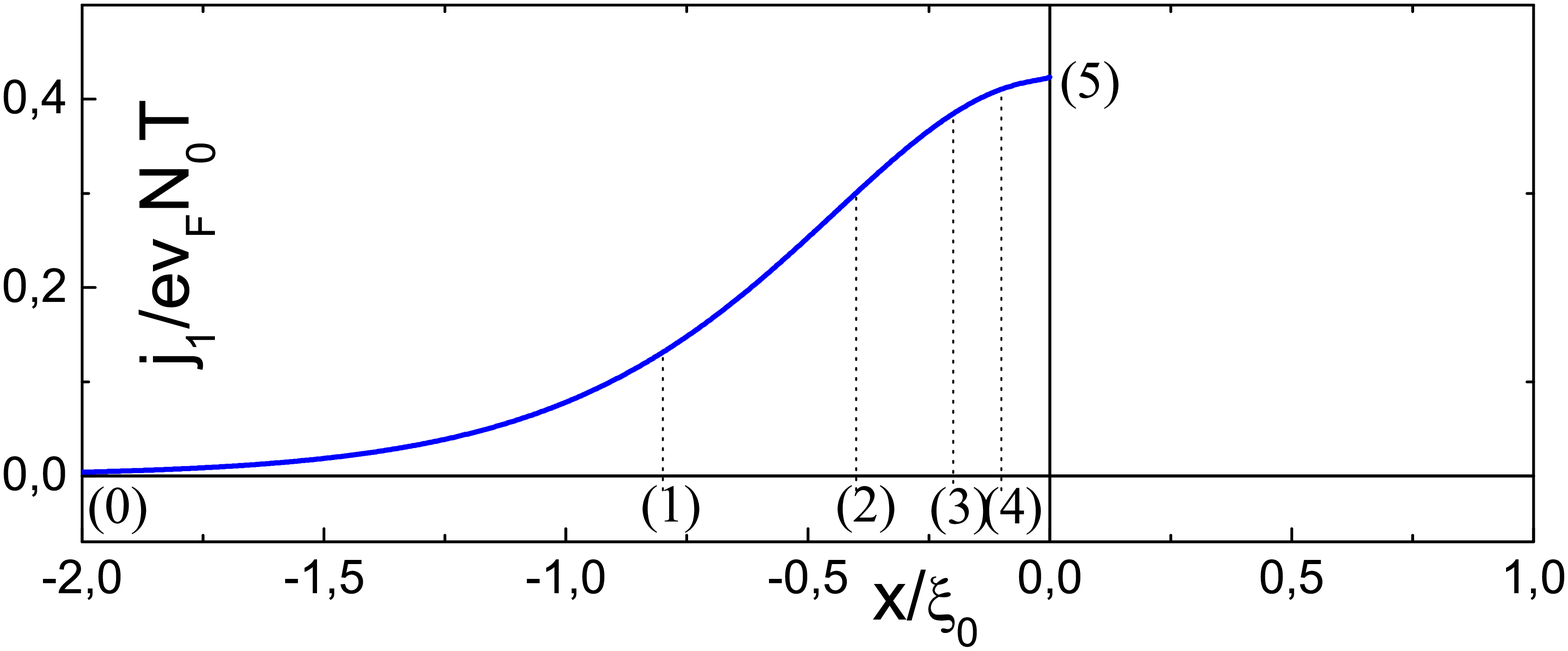}} \\
\vspace{-2 mm}
\raggedright{b)}
\end{minipage}
\end{minipage}
\hfill
\begin{minipage}[h]{0.45\linewidth}
\center{\includegraphics[width=1\linewidth]{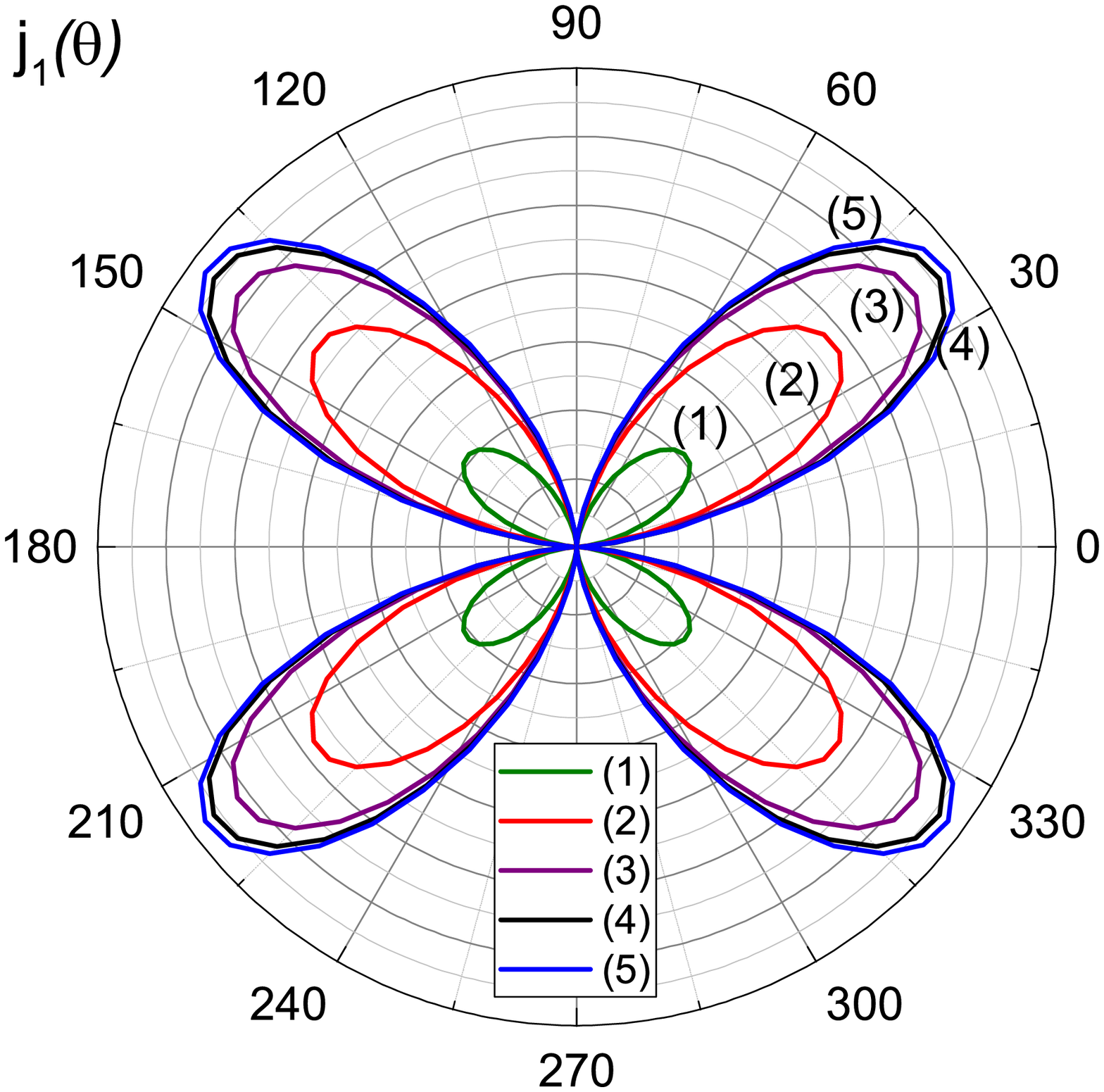}} \\
\vspace{-8 mm}
\raggedright{c)}
\end{minipage}
\vfill
\begin{minipage}[h]{\linewidth}
\begin{minipage}[h]{0.3\linewidth}
\center{\includegraphics[width=1\linewidth]{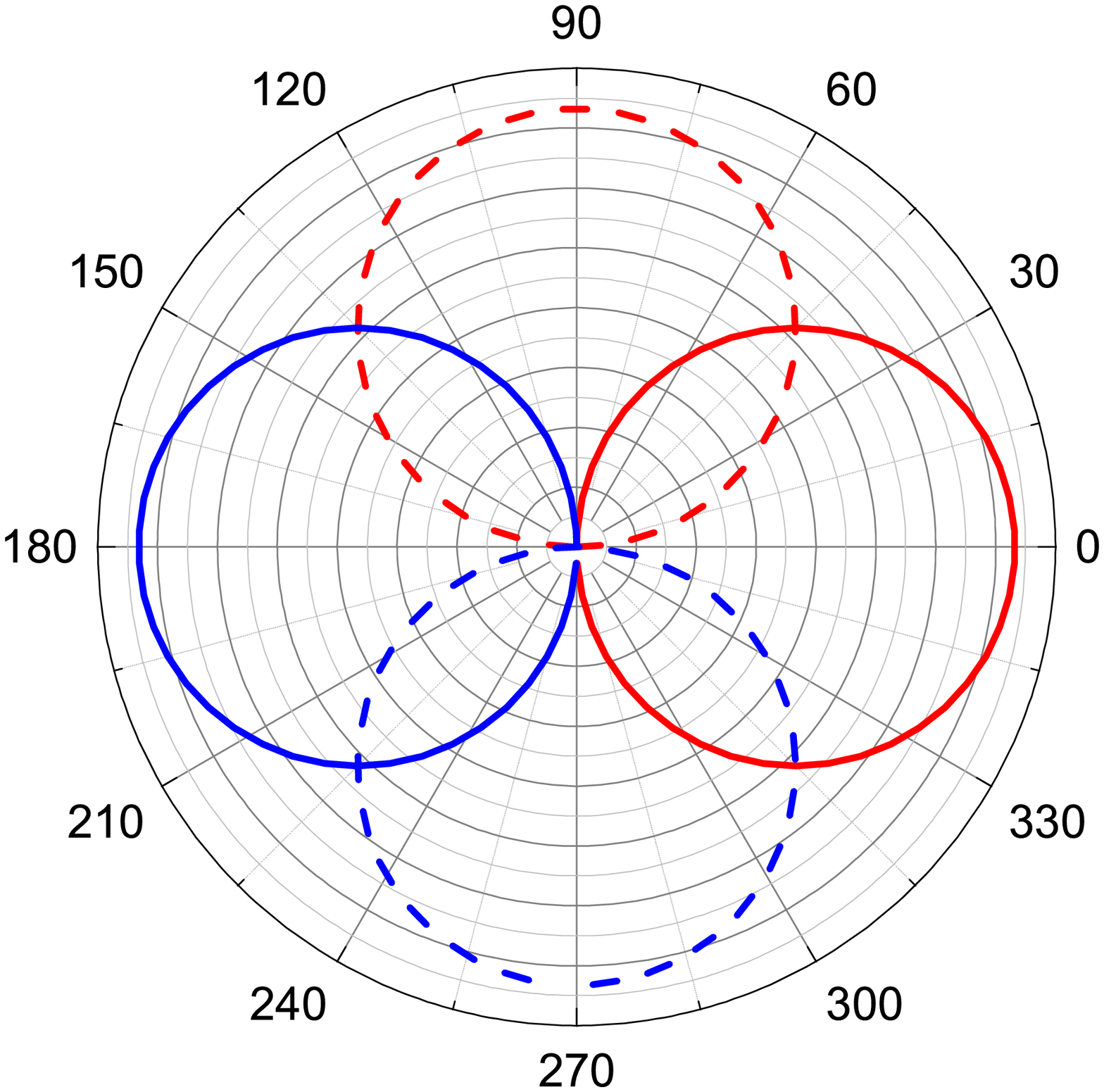}} \\
\vspace{-6 mm}
\raggedright{\small{(0)}}
\end{minipage}
\hfill
\begin{minipage}[h]{0.3\linewidth}
\center{\includegraphics[width=1\linewidth]{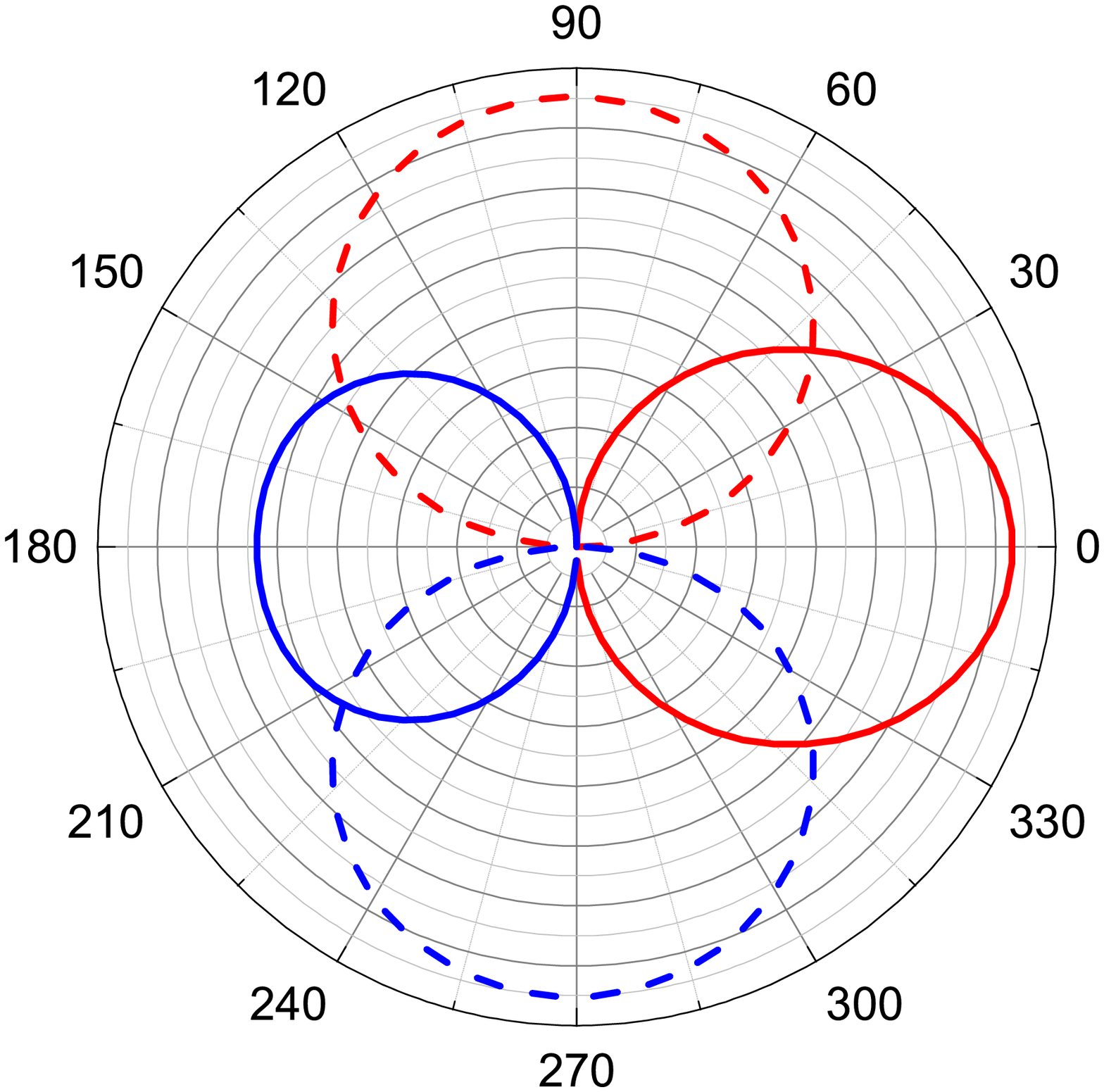}} \\
\vspace{-6 mm}
\raggedright{\small{(1)}}
\end{minipage}
\hfill
\begin{minipage}[h]{0.3\linewidth}
\center{\includegraphics[width=1\linewidth]{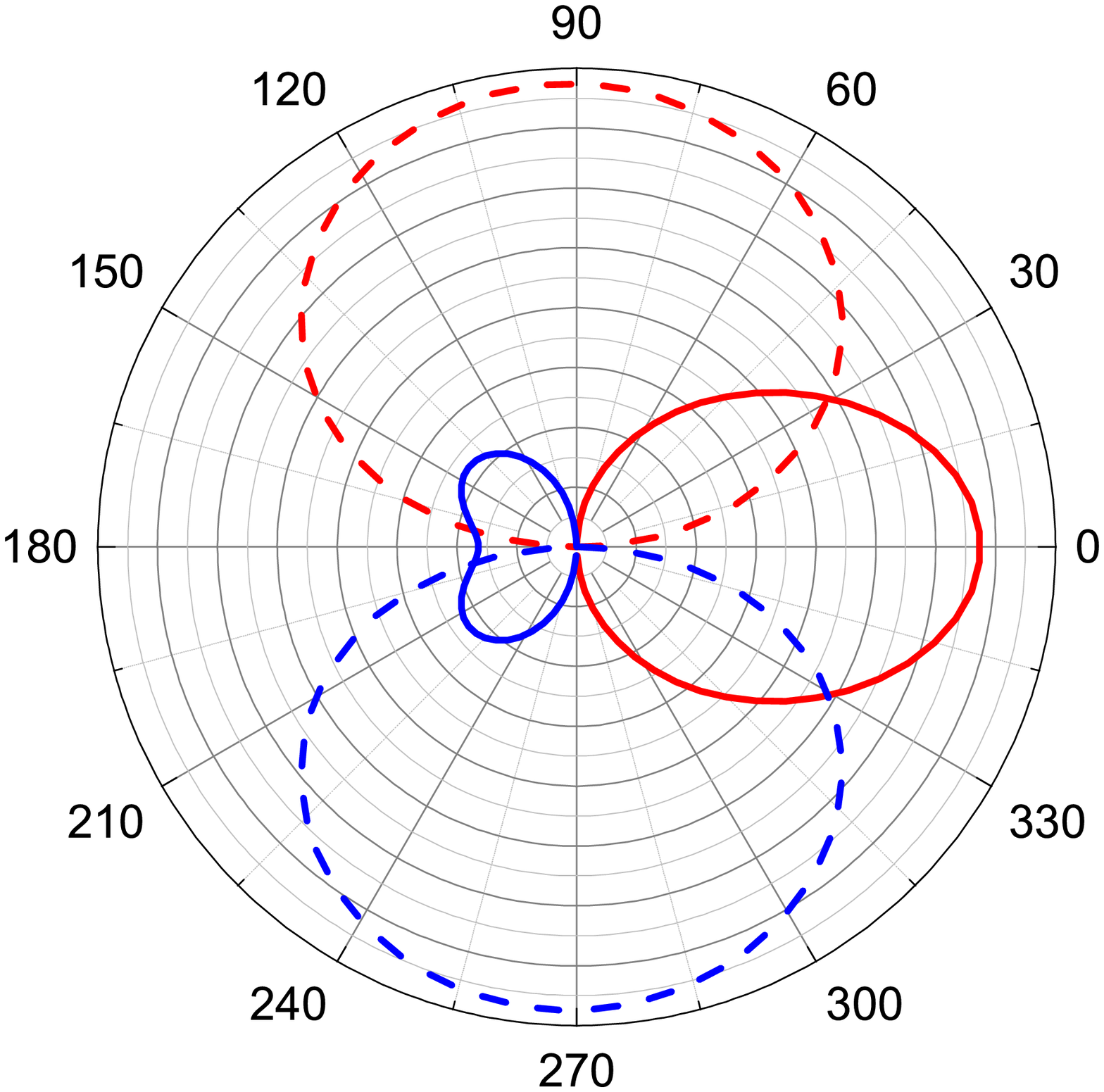}} \\
\vspace{-6 mm}
\raggedright{\small{(2)}}
\end{minipage}
\vfill
\begin{minipage}[h]{0.3\linewidth}
\center{\includegraphics[width=1\linewidth]{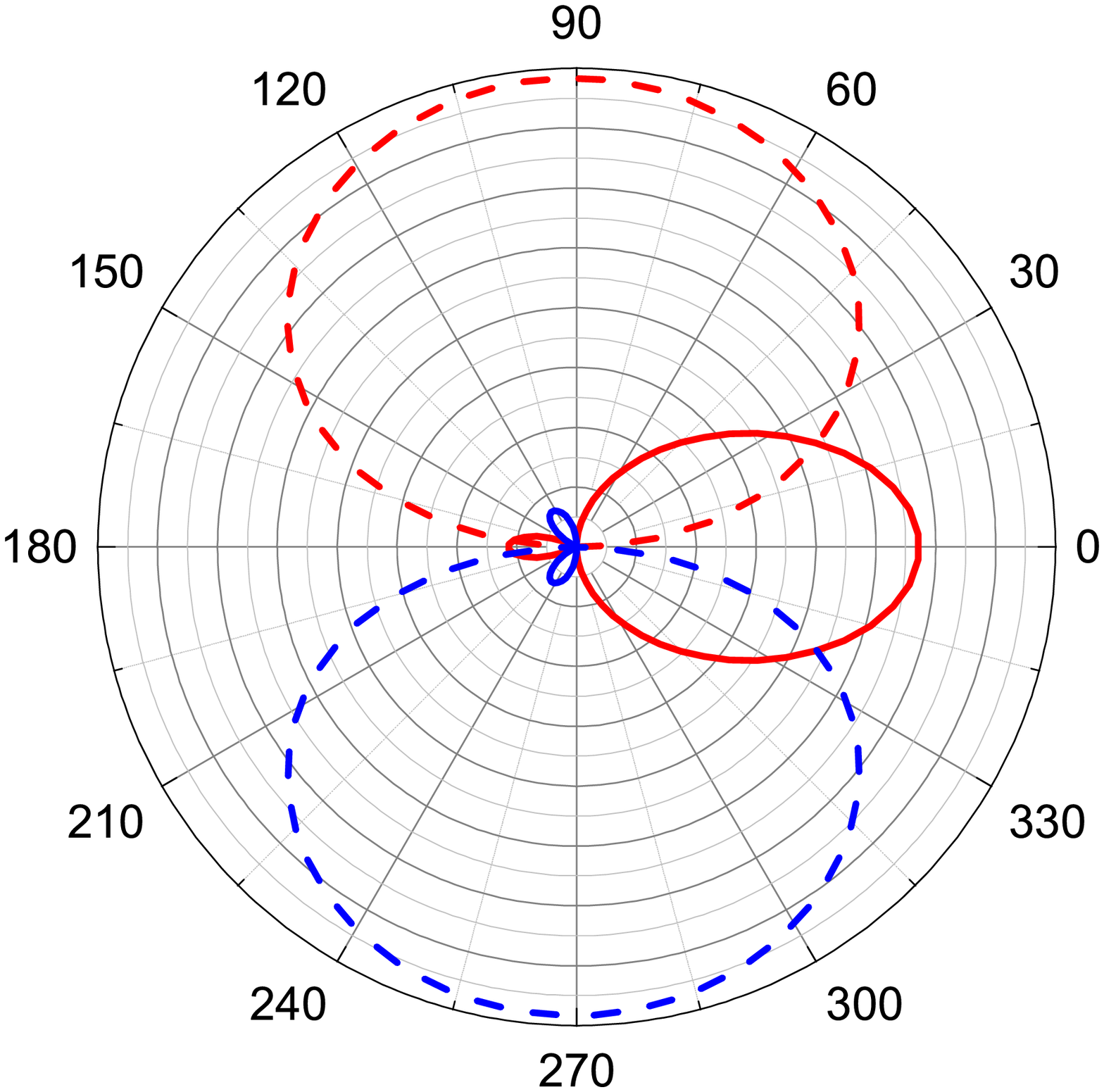}} \\
\vspace{-6 mm}
\raggedright{\small{(3)}}
\end{minipage}
\hfill
\begin{minipage}[h]{0.3\linewidth}
\center{\includegraphics[width=1\linewidth]{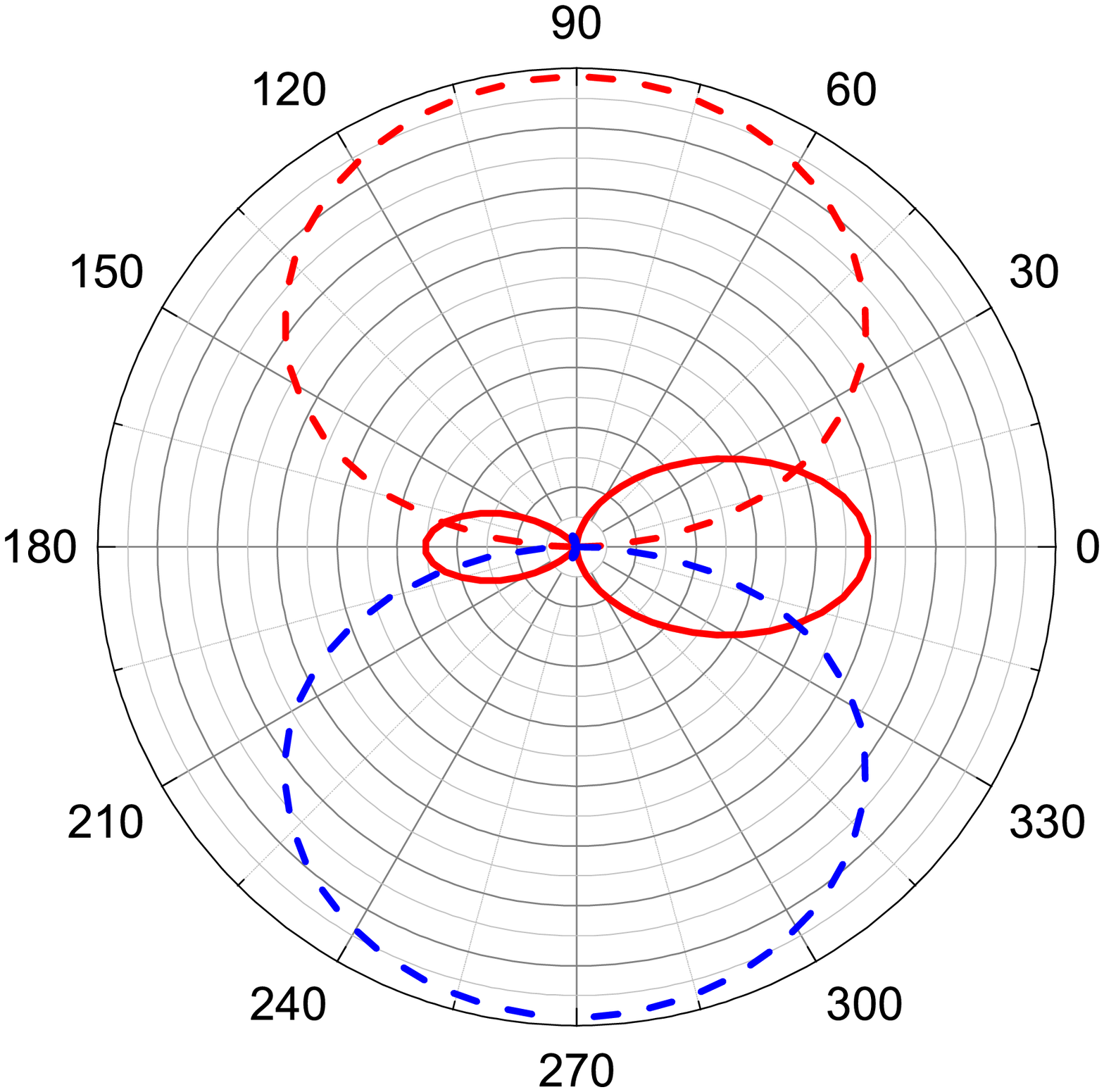}} \\
\vspace{-6 mm}
\raggedright{\small{(4)}}
\end{minipage}
\hfill
\begin{minipage}[h]{0.3\linewidth}
\center{\includegraphics[width=1\linewidth]{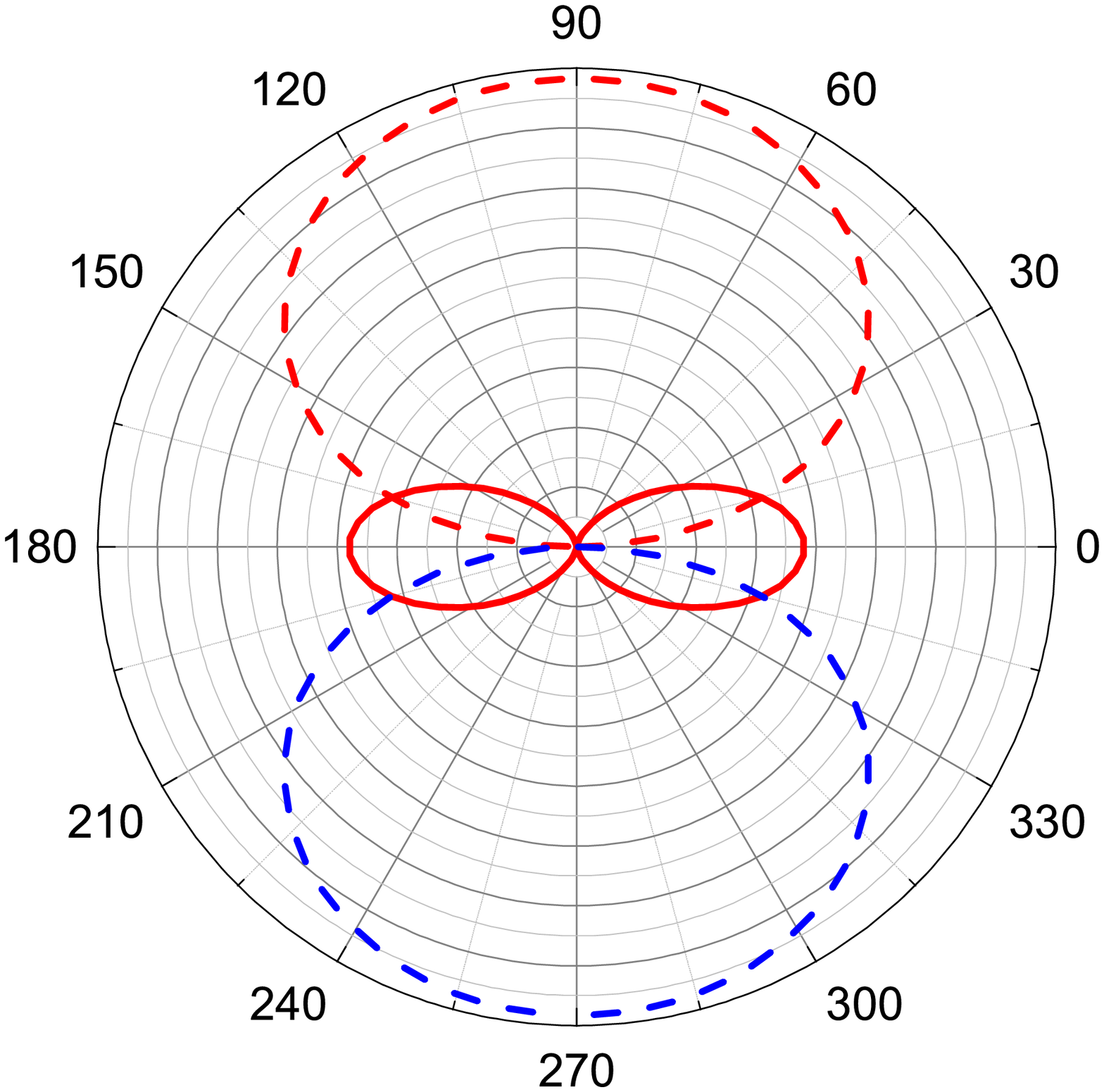}} \\
\vspace{-6 mm}
\raggedright{\small{(5)}}
\end{minipage}
\end{minipage}
\raggedright{d)}
\caption{(Color Online) Spatial distribution of set a) - d) of parameters in
p-wave superconductor with specular surface: \newline
a) the pair potentials $\protect\Delta_x$ and $\protect\Delta_y$ and b) the
surface current density $j_1$, c) contribution of particles with angle $%
\protect\theta$ in the formation of surface current $j_1$ at the different
points of the structure, d) angle dependent pair amplitude $f_1(\protect%
\theta)$ at the different points (0)-(5) of the structure. At the panel d)
solid lines correspond to the real part of $f$ and dashed lines correspond
to the imaginary part of $f $; red color means positive value and blue color
means negative one. Set of the points is following: (0) $x = -2 \protect\xi%
_0 $, (1) $x = -0.8 \protect\xi_0 $, (2) $-0.4 \protect\xi_0 $, (3) $-0.2
\protect\xi_0 $, (4) $-0.1 \protect\xi_0 $, (5) - $x = 0 \protect\xi_0 $.
All calculations were performed at $T=0.5 T_C$ and $d=0$.}
\label{F_Specular}
\end{figure*}

Contrary to that, in the considered situation vanishing of $\Delta_x$ order parameter component is not accompanied by vanishing of anomalous functions at a specular surface, see the plot Fig. \ref{F_Specular}d(5).  Moreover, due to the conditions (\ref{BC1a}), (\ref{BC1b}) the sign of the $f_1 (\theta)$ functions on outgoing directories is opposite to that in the bulk. It is exactly the mechanism that results in nucleation of spontaneous current at free mirror surface.  Figure 1 gives that normalized amplitude of the first item in the expression for the supercurrent is slightly larger that $0.4.$

\begin{figure*}[h]
\begin{minipage}[h]{0.5\linewidth}
\begin{minipage}[h]{0.99\linewidth}
\center{\includegraphics[width=1\linewidth]{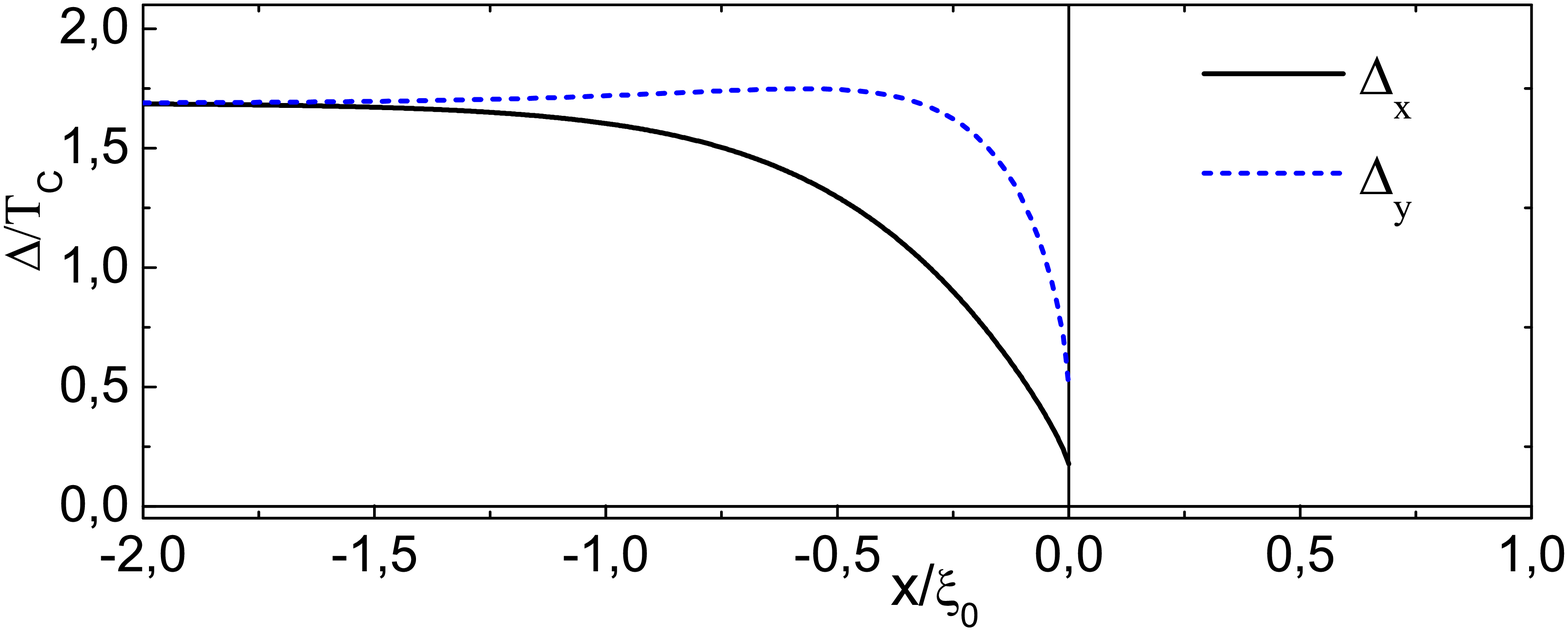}} \\
\vspace{-2 mm}
\raggedright{a)}
\end{minipage}
\vfill
\begin{minipage}[h]{0.99\linewidth}
\center{\includegraphics[width=1\linewidth]{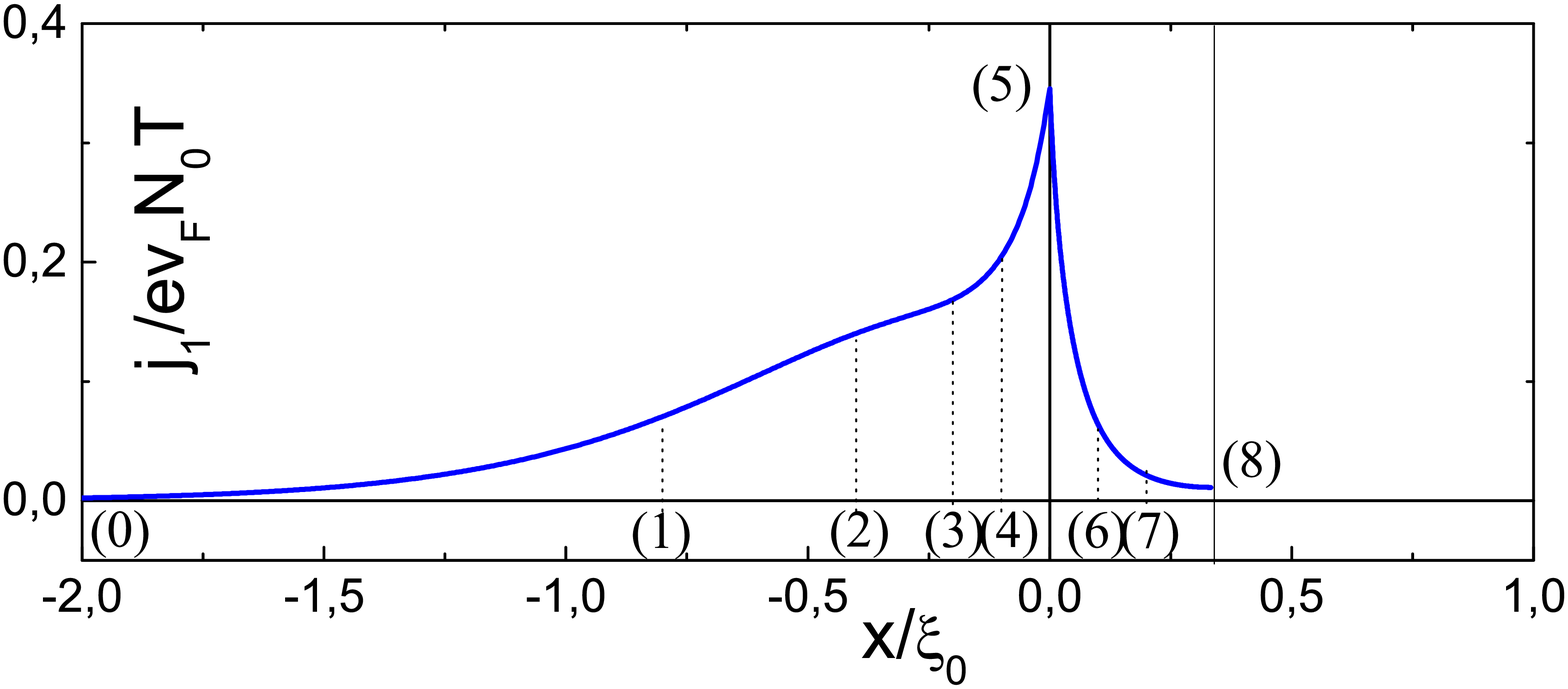}} \\
\vspace{-2 mm}
\raggedright{b)}
\end{minipage}
\end{minipage}
\hfill
\begin{minipage}[h]{0.45\linewidth}
\center{\includegraphics[width=1\linewidth]{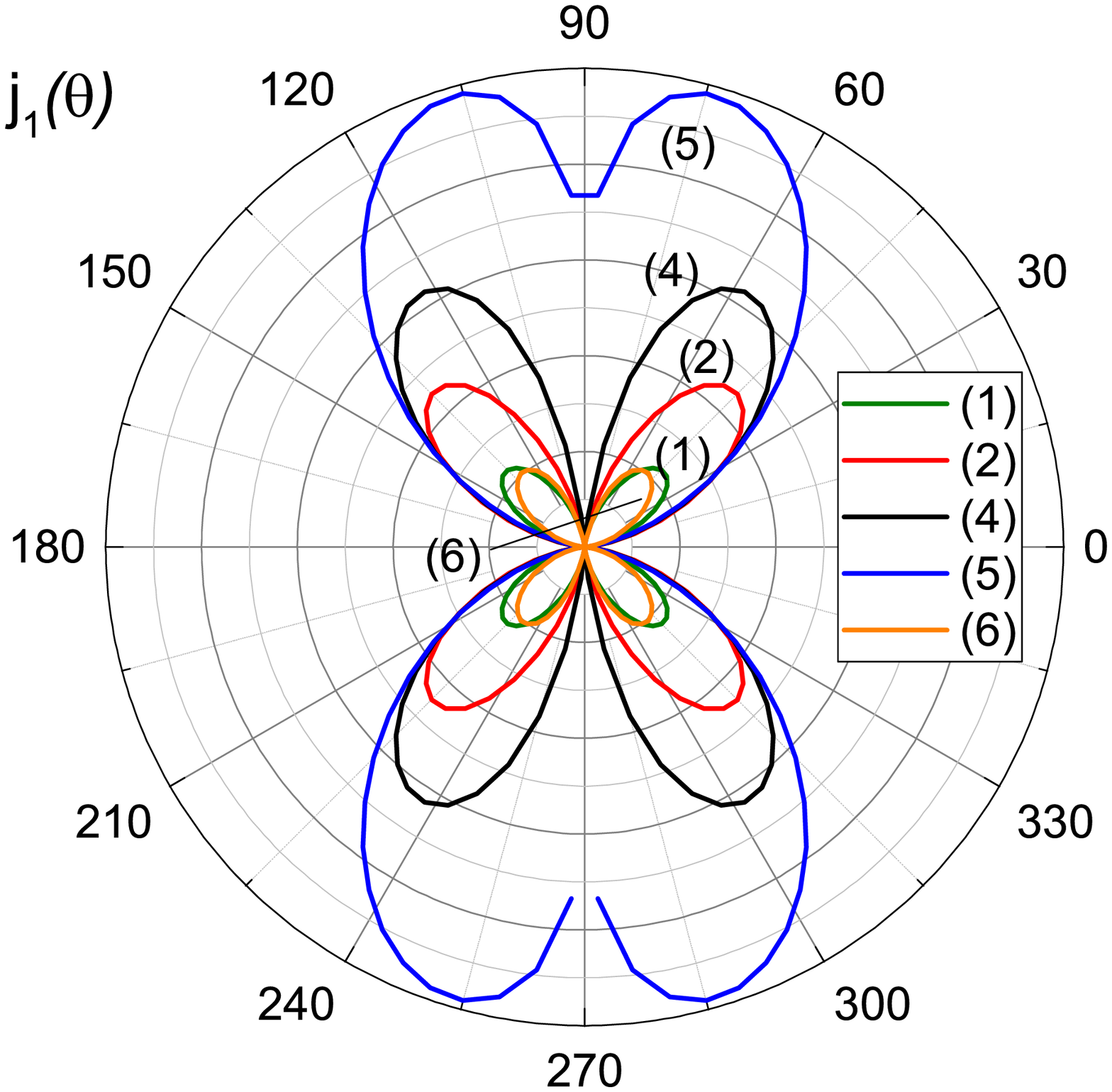}} \\
\vspace{-8 mm}
\raggedright{c)}
\end{minipage}
\vfill
\begin{minipage}[h]{\linewidth}
\begin{minipage}[h]{0.3\linewidth}
\center{\includegraphics[width=1\linewidth]{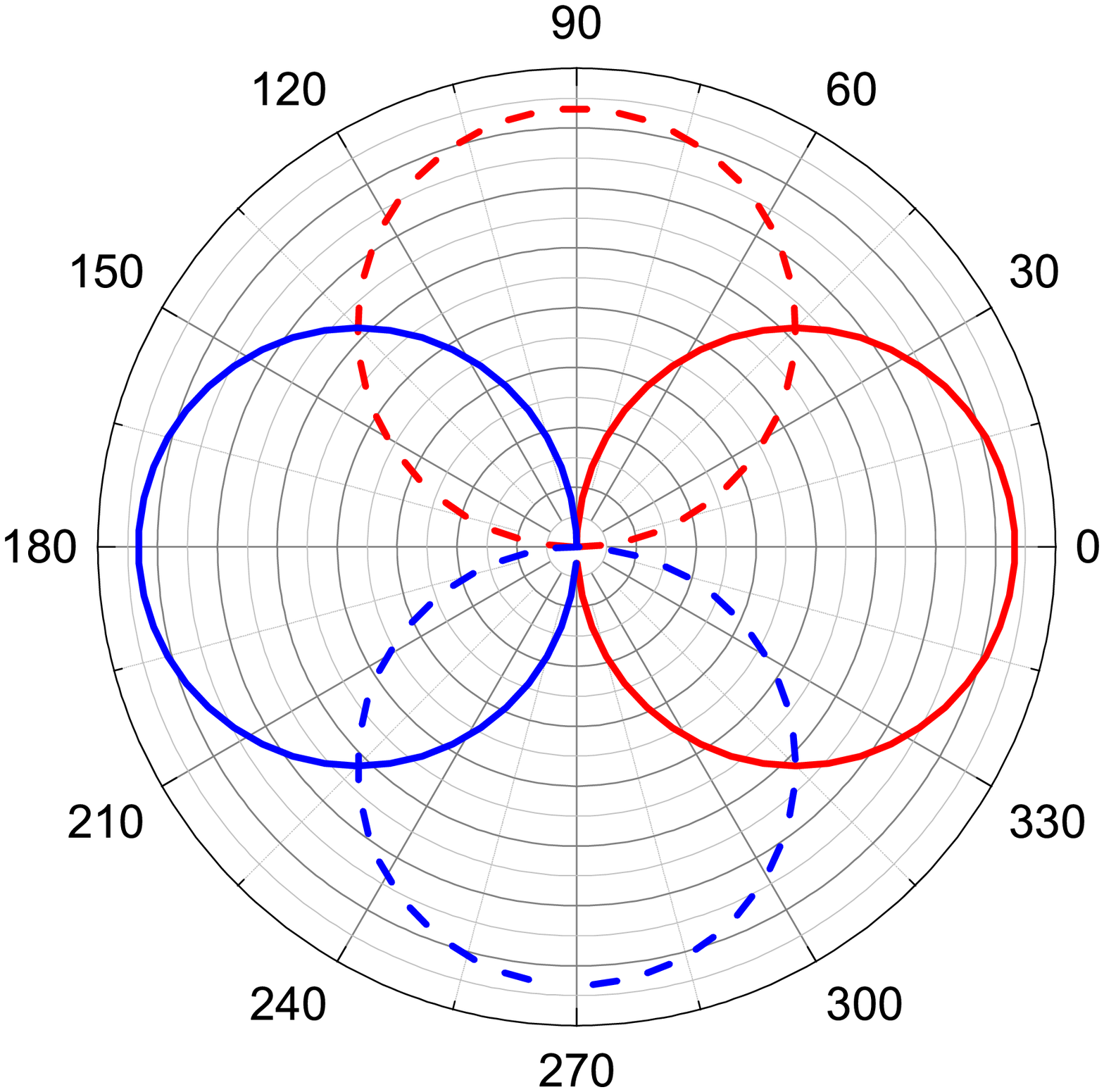}} \\
\vspace{-6 mm}
\raggedright{\small{(0)}}
\end{minipage}
\hfill
\begin{minipage}[h]{0.3\linewidth}
\center{\includegraphics[width=1\linewidth]{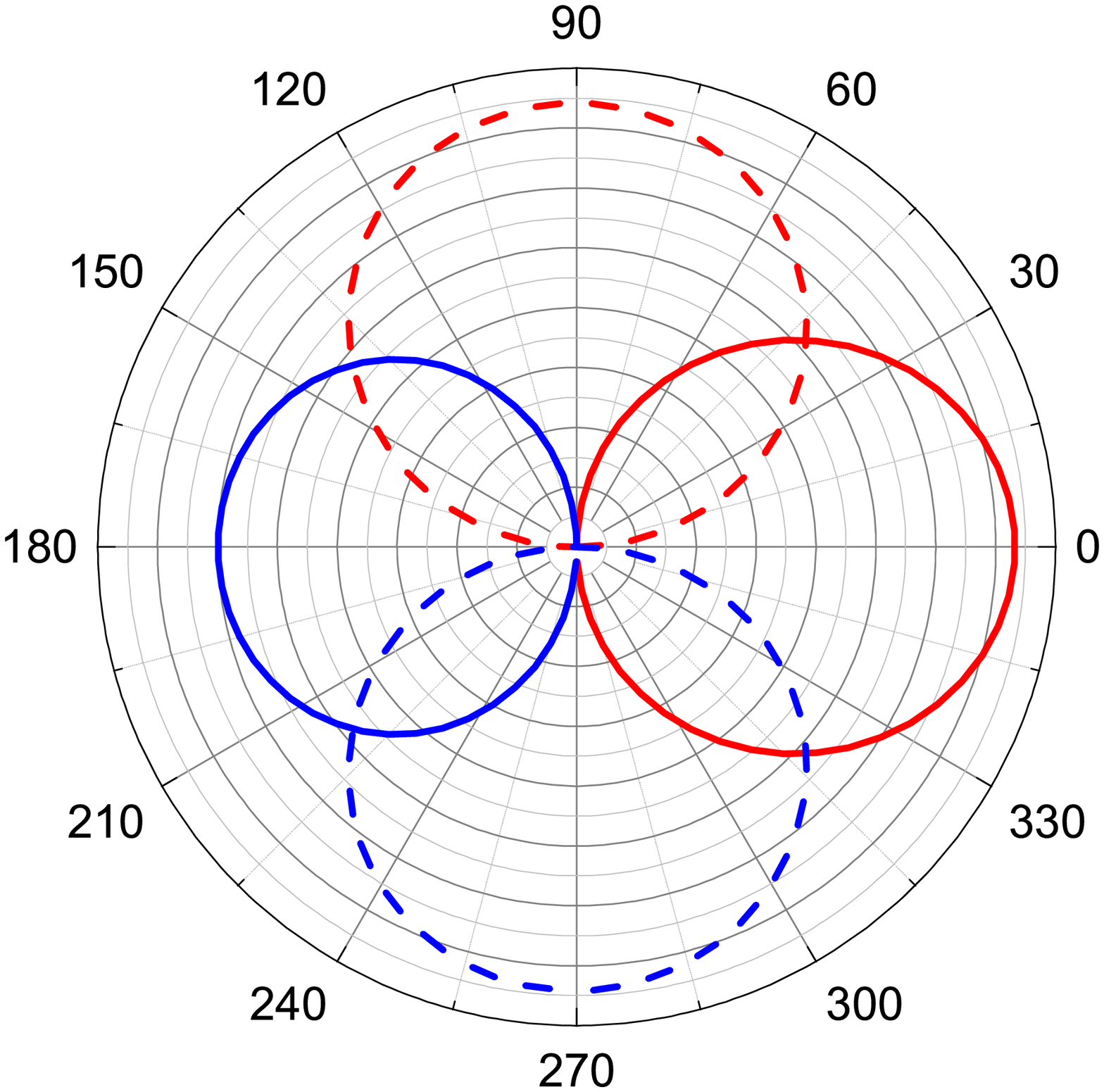}} \\
\vspace{-6 mm}
\raggedright{\small{(1)}}
\end{minipage}
\hfill
\begin{minipage}[h]{0.3\linewidth}
\center{\includegraphics[width=1\linewidth]{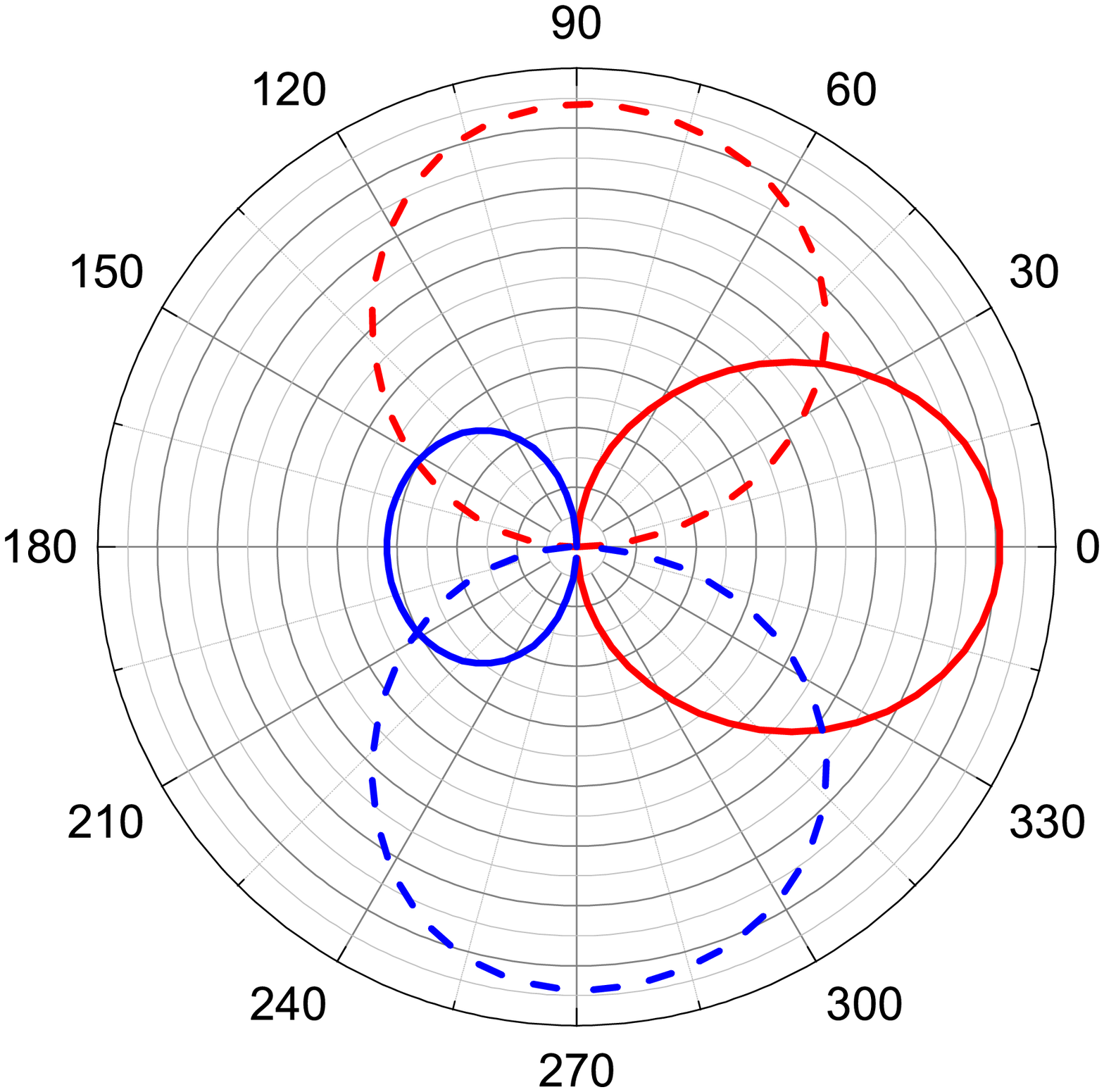}} \\
\vspace{-6 mm}
\raggedright{\small{(2)}}
\end{minipage}
\vfill
\begin{minipage}[h]{0.3\linewidth}
\center{\includegraphics[width=1\linewidth]{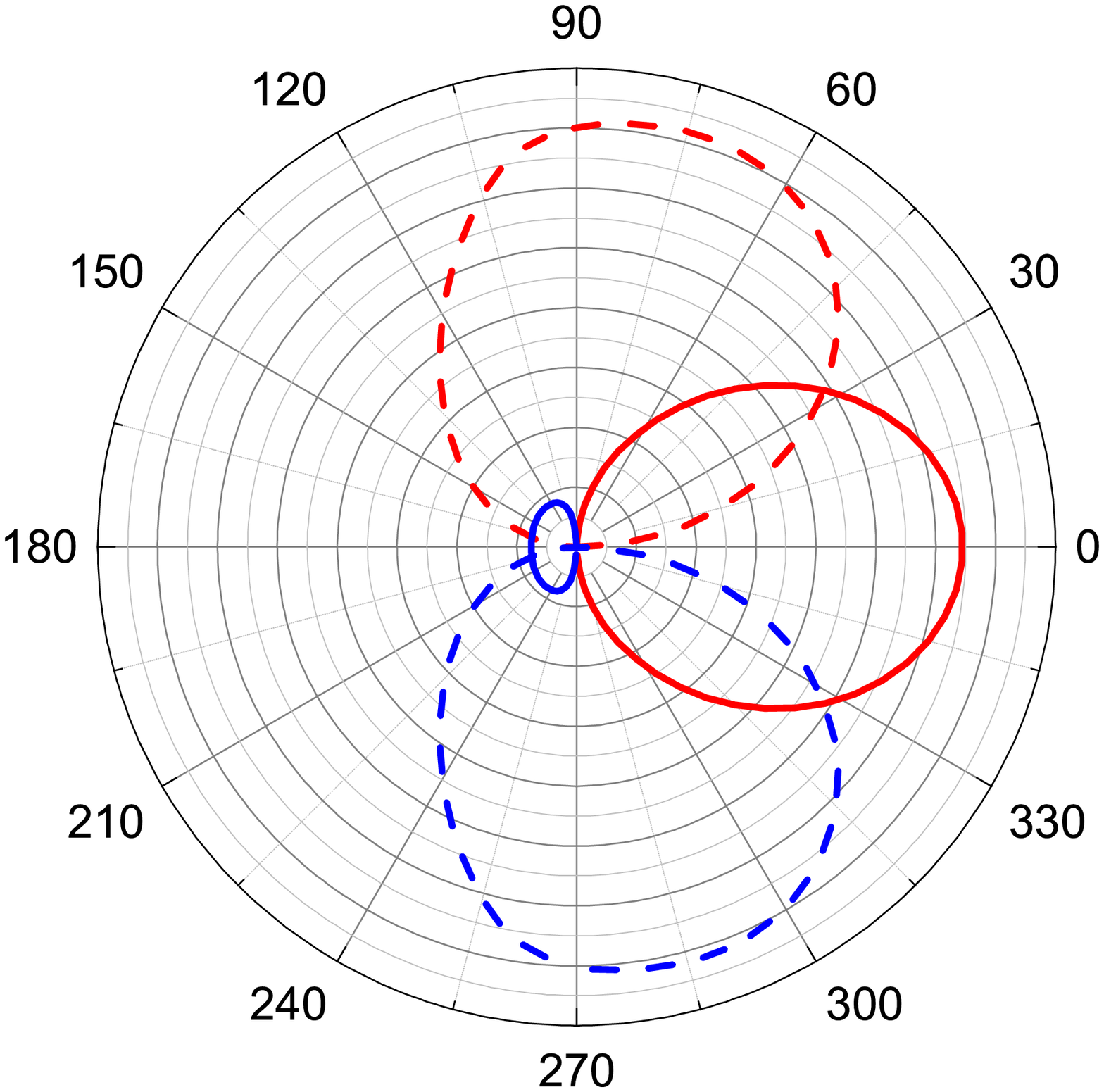}} \\
\vspace{-6 mm}
\raggedright{\small{(3)}}
\end{minipage}
\hfill
\begin{minipage}[h]{0.3\linewidth}
\center{\includegraphics[width=1\linewidth]{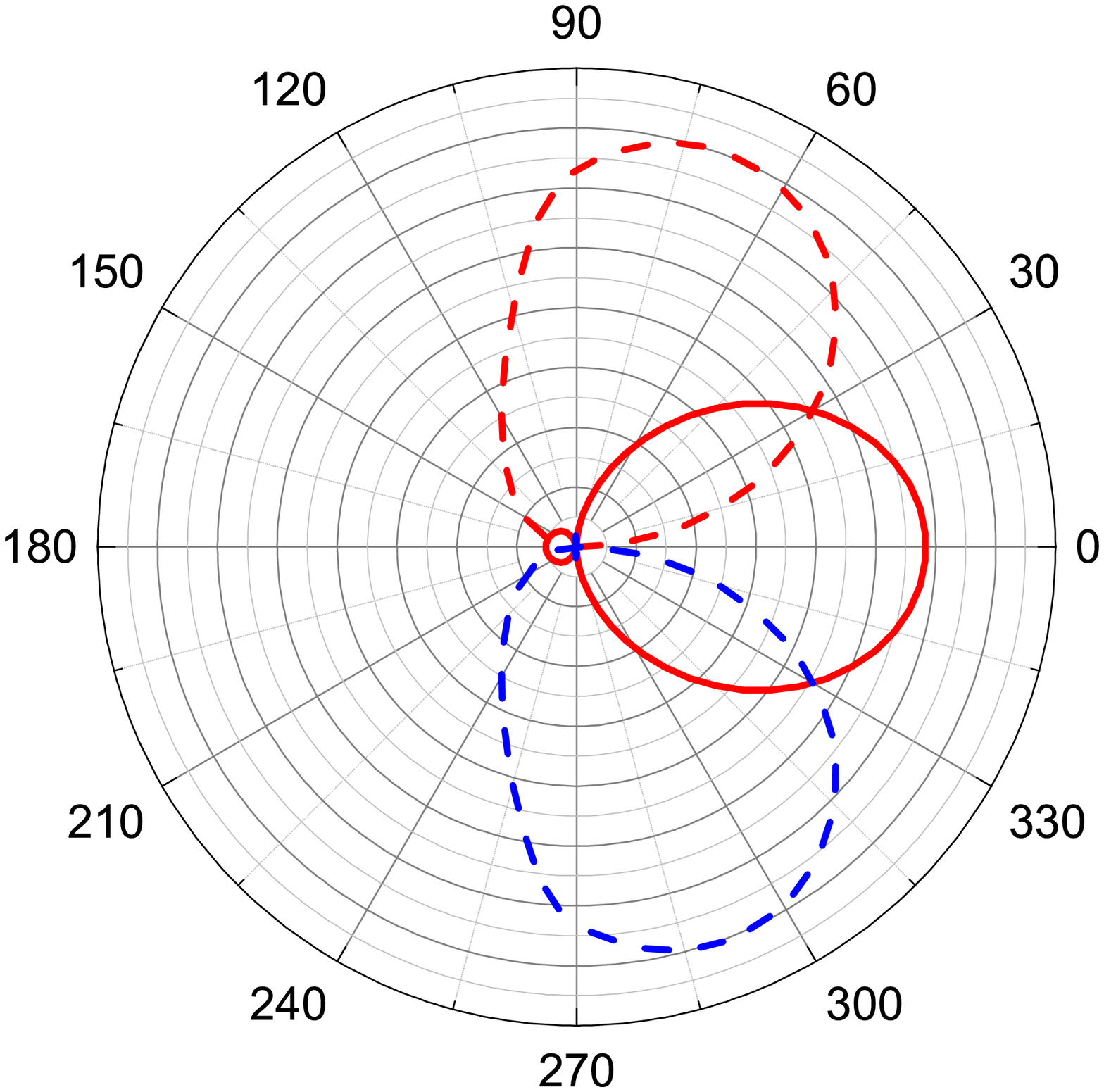}} \\
\vspace{-6 mm}
\raggedright{\small{(4)}}
\end{minipage}
\hfill
\begin{minipage}[h]{0.3\linewidth}
\center{\includegraphics[width=1\linewidth]{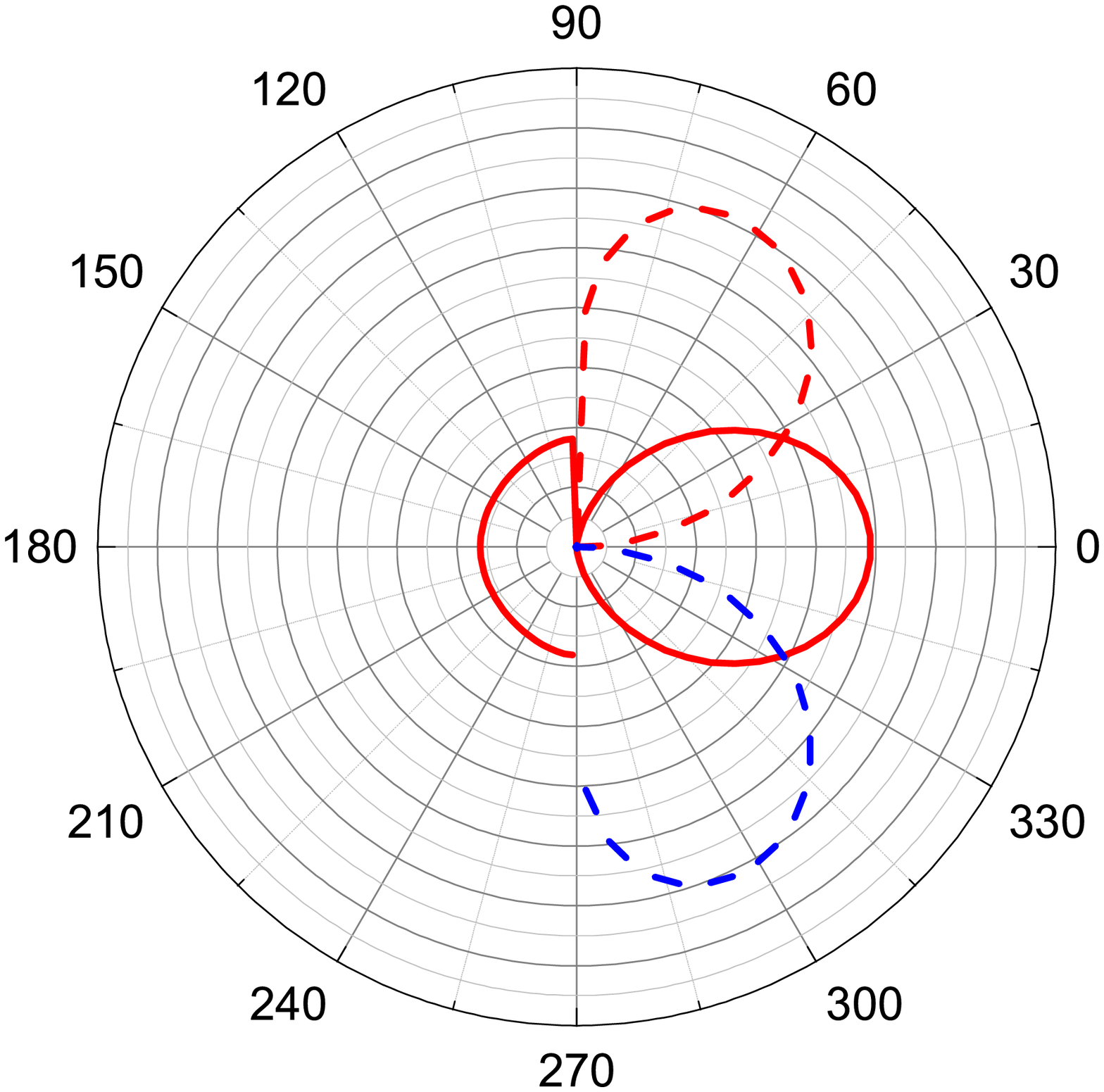}} \\
\vspace{-6 mm}
\raggedright{\small{(5)}}
\end{minipage}
\vfill
\begin{minipage}[h]{0.3\linewidth}
\center{\includegraphics[width=1\linewidth]{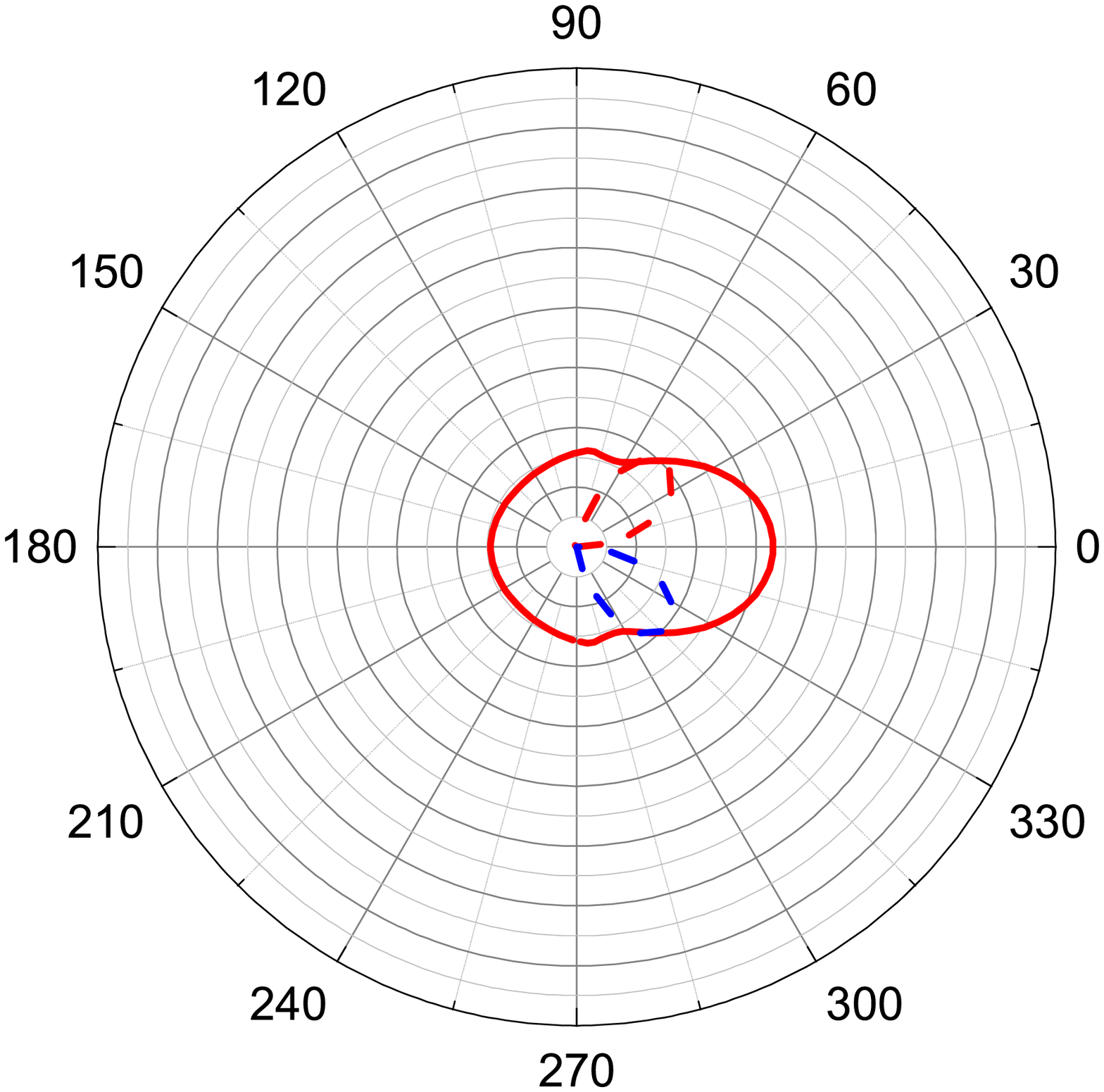}} \\
\vspace{-6 mm}
\raggedright{\small{(6)}}
\end{minipage}
\hfill
\begin{minipage}[h]{0.3\linewidth}
\center{\includegraphics[width=1\linewidth]{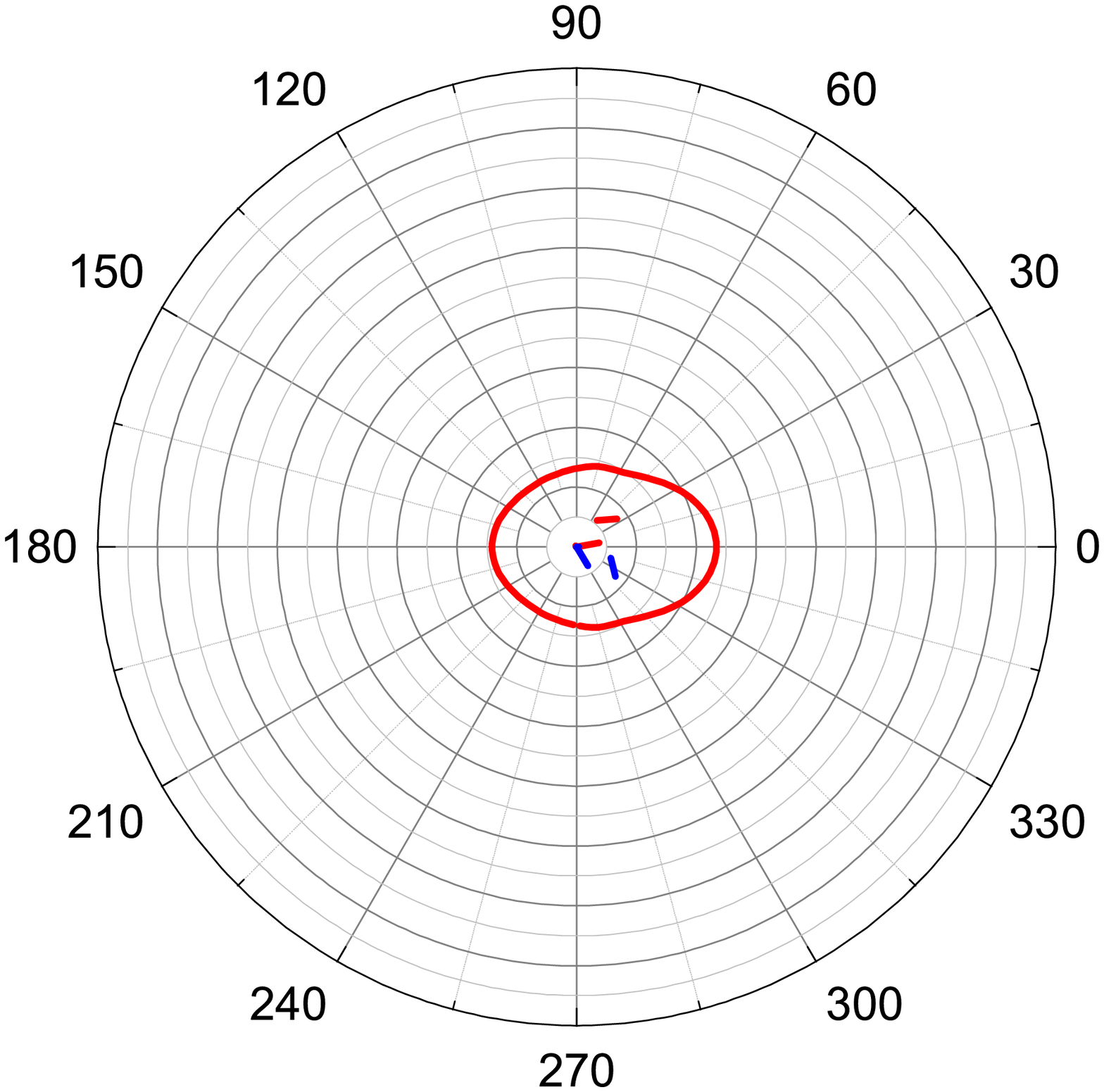}} \\
\vspace{-6 mm}
\raggedright{\small{(7)}}
\end{minipage}
\hfill
\begin{minipage}[h]{0.3\linewidth}
\center{\includegraphics[width=1\linewidth]{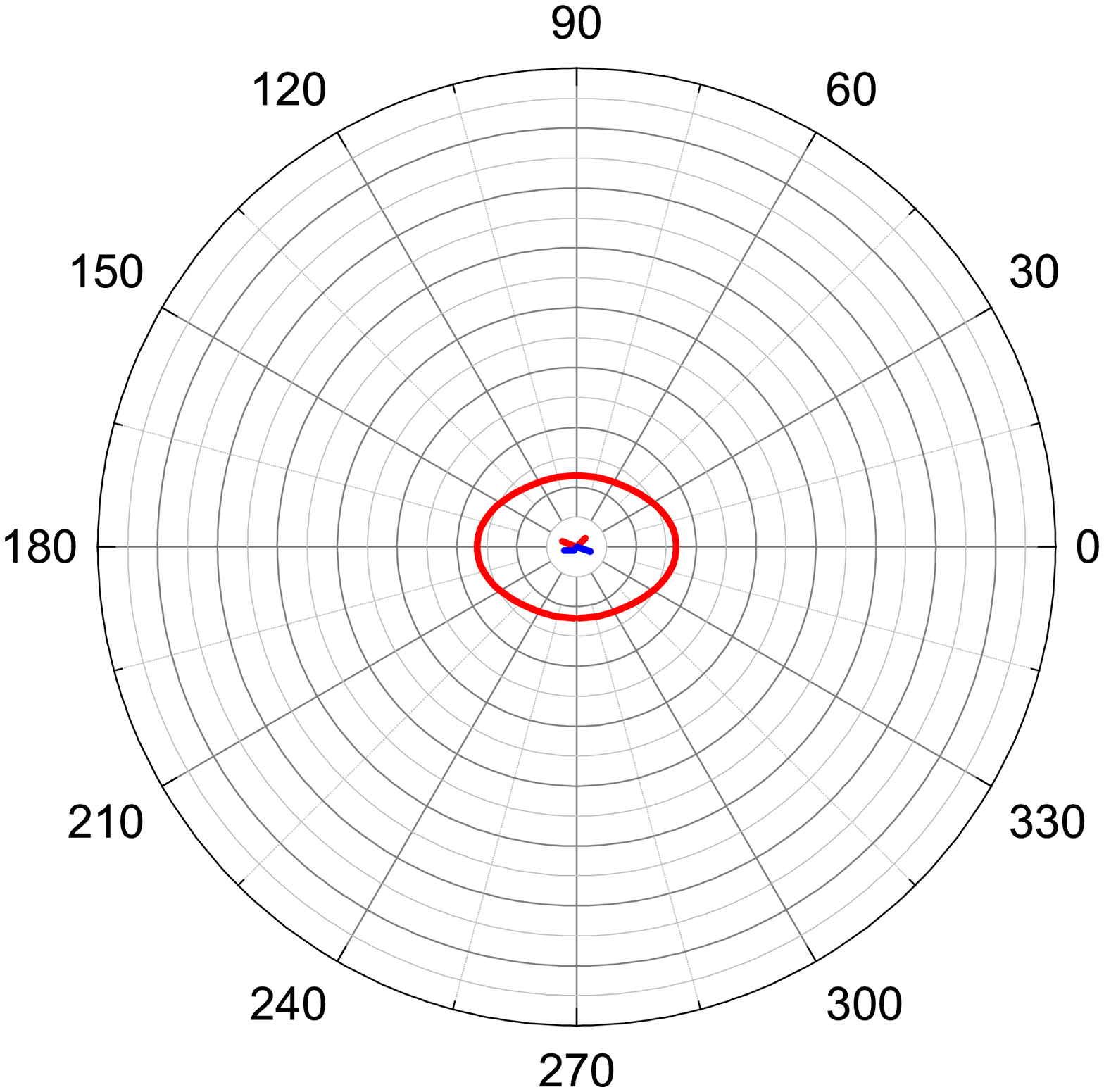}} \\
\vspace{-6 mm}
\raggedright{\small{(8)}}
\end{minipage}
\end{minipage}
\raggedright{d)}
\caption{(Color Online) Spatial distribution of set a) - d) of parameters in
p-wave superconductor with rough surface: \newline
a) the pair potentials $\protect\Delta_x$ and $\protect\Delta_y$ and b) the
surface current density $j_1$, c) contribution of particles with angle $%
\protect\theta$ in the formation of surface current $j_1$ at the different
points of the structure, d) angle dependent pair amplitude $f_1(\protect%
\theta)$ at the different points (0)-(8) of the structure. Set of the points
is following: (0) $x = -2 \protect\xi_0 $, (1) $-0.8 \protect\xi_0 $, (2) $%
-0.4 \protect\xi_0 $, (3) $-0.2 \protect\xi_0 $, (4) $-0.1 \protect\xi_0 $,
(5) $0 \protect\xi_0 $, (6) $0.1 \protect\xi_0 $, (7) $0.2 \protect\xi_0 $,
(8) the surface $x = 0.33 \protect\xi_0 $. All calculations were performed
at $T=0.5 T_C$, $d= 1 l_e$ and $\protect\xi_0=3 l_e$ .}
\label{f_rough}
\end{figure*}

\begin{figure*}[h]
\begin{minipage}[h]{0.5\linewidth}
\begin{minipage}[h]{0.99\linewidth}
\center{\includegraphics[width=1\linewidth]{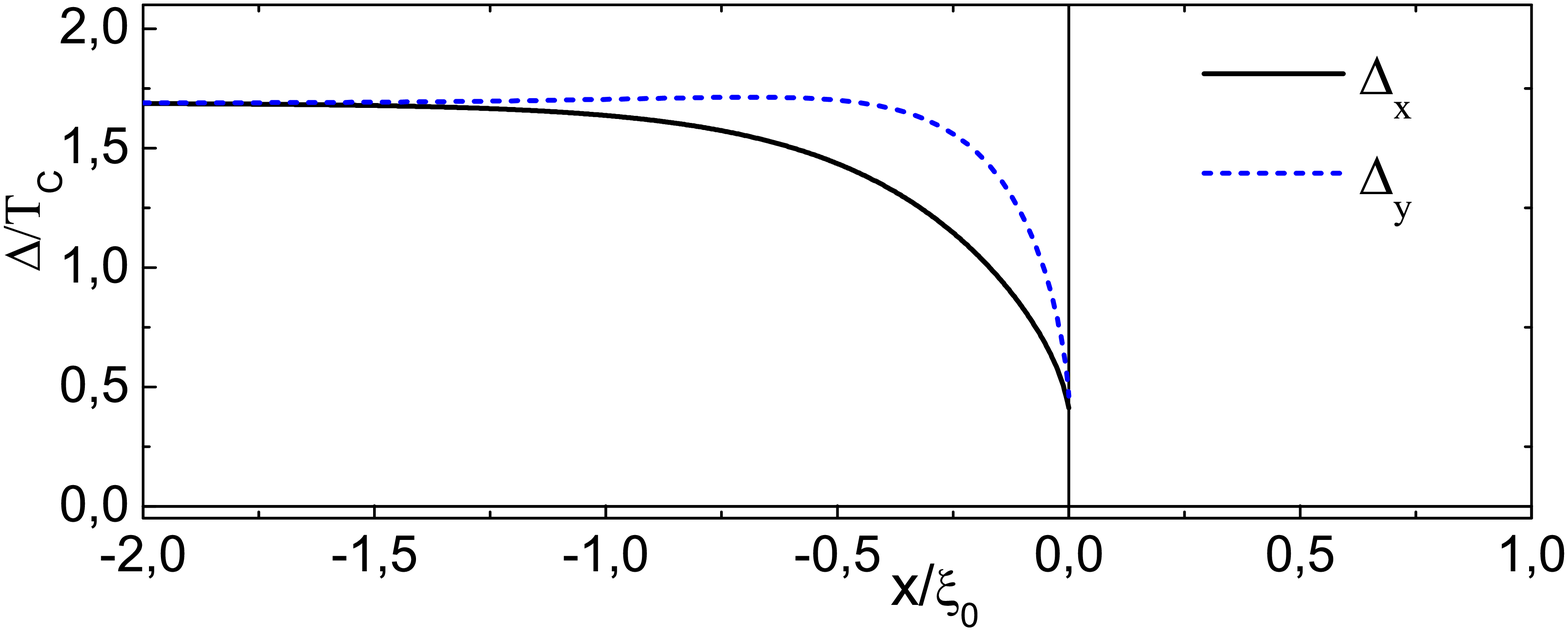}} \\
\vspace{-2 mm}
\raggedright{a)}
\end{minipage}
\vfill
\begin{minipage}[h]{0.99\linewidth}
\center{\includegraphics[width=1\linewidth]{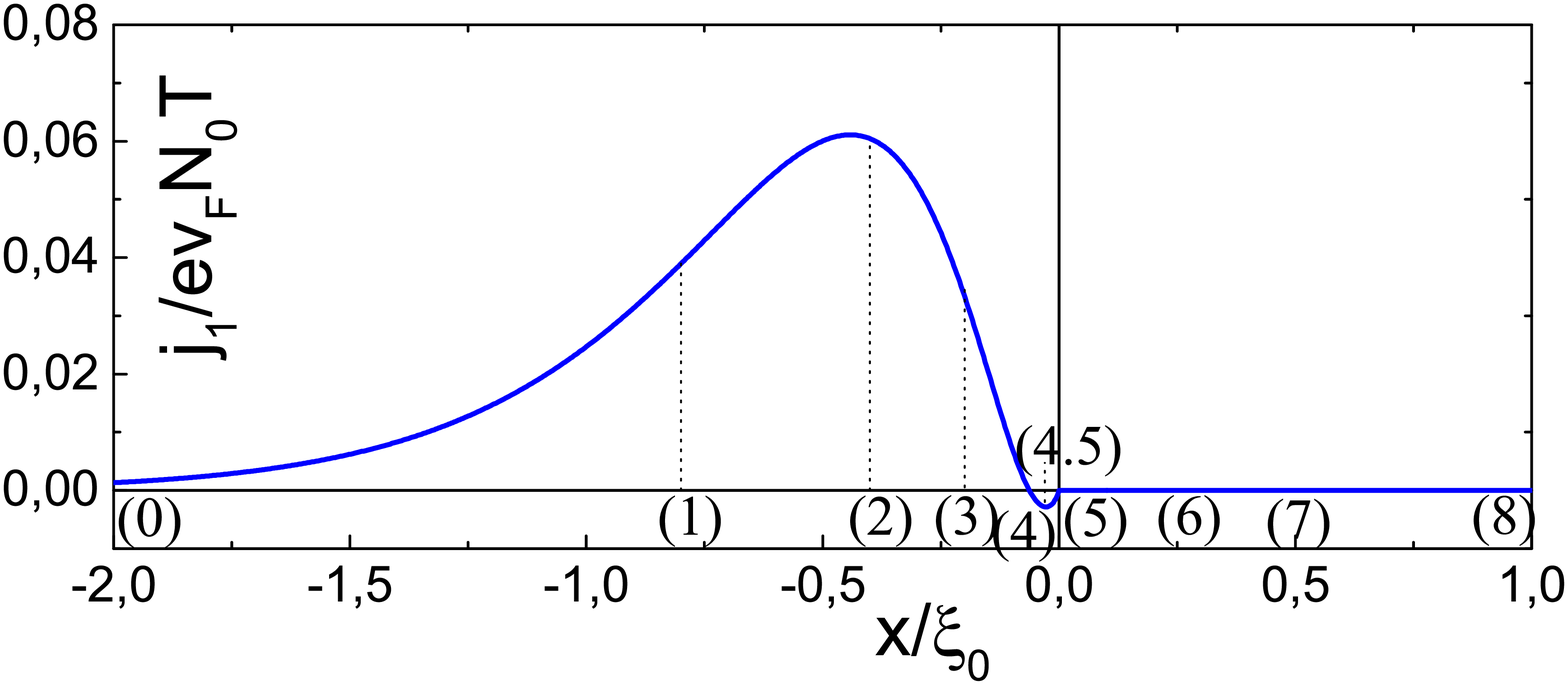}} \\
\vspace{-2 mm}
\raggedright{b)}
\end{minipage}
\end{minipage}
\hfill
\begin{minipage}[h]{0.45\linewidth}
\center{\includegraphics[width=1\linewidth]{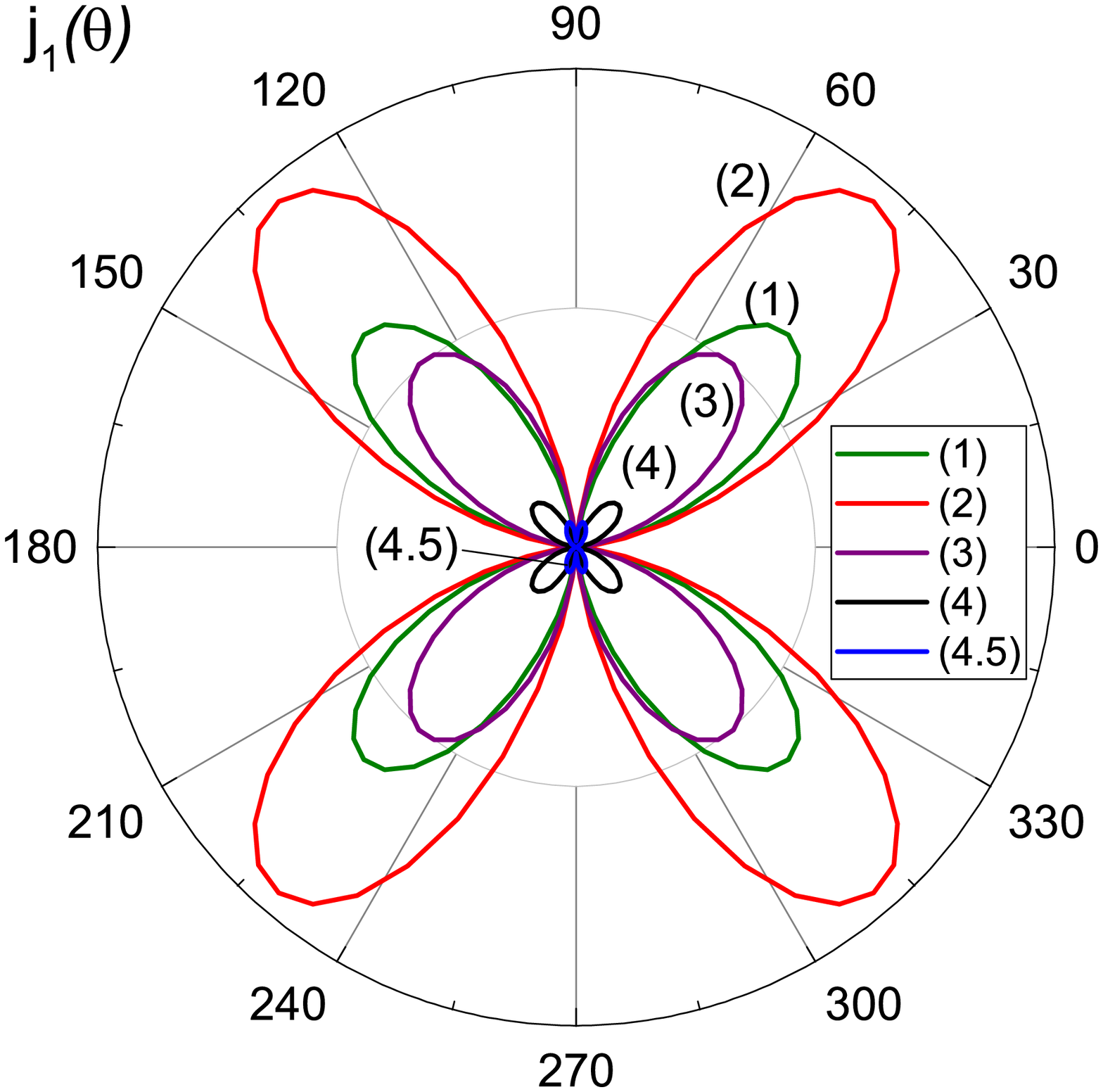}} \\
\vspace{-8 mm}
\raggedright{c)}
\end{minipage}
\vfill
\begin{minipage}[h]{\linewidth}
\begin{minipage}[h]{0.3\linewidth}
\center{\includegraphics[width=1\linewidth]{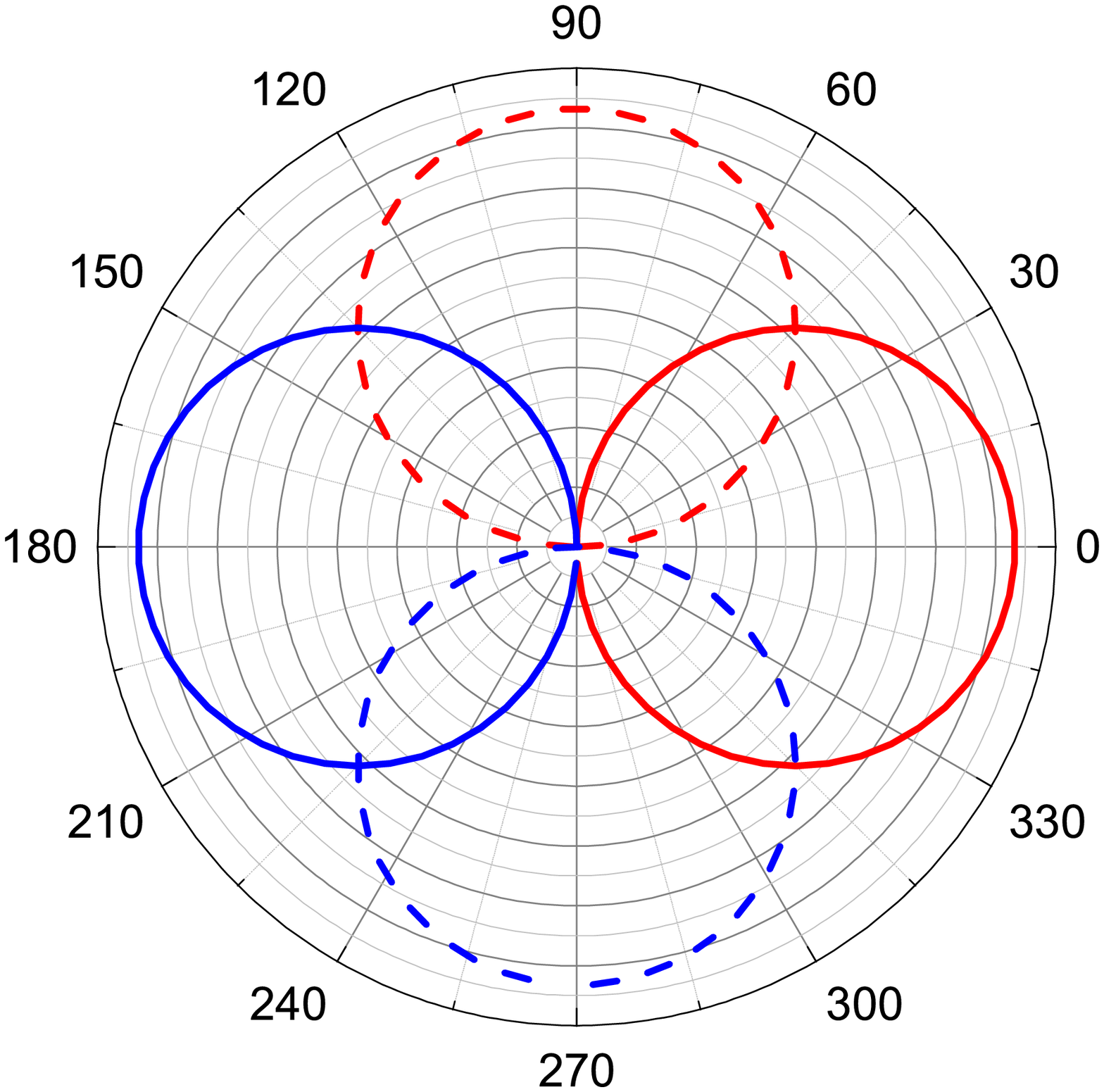}} \\
\vspace{-6 mm}
\raggedright{\small{(0)}}
\end{minipage}
\hfill
\begin{minipage}[h]{0.3\linewidth}
\center{\includegraphics[width=1\linewidth]{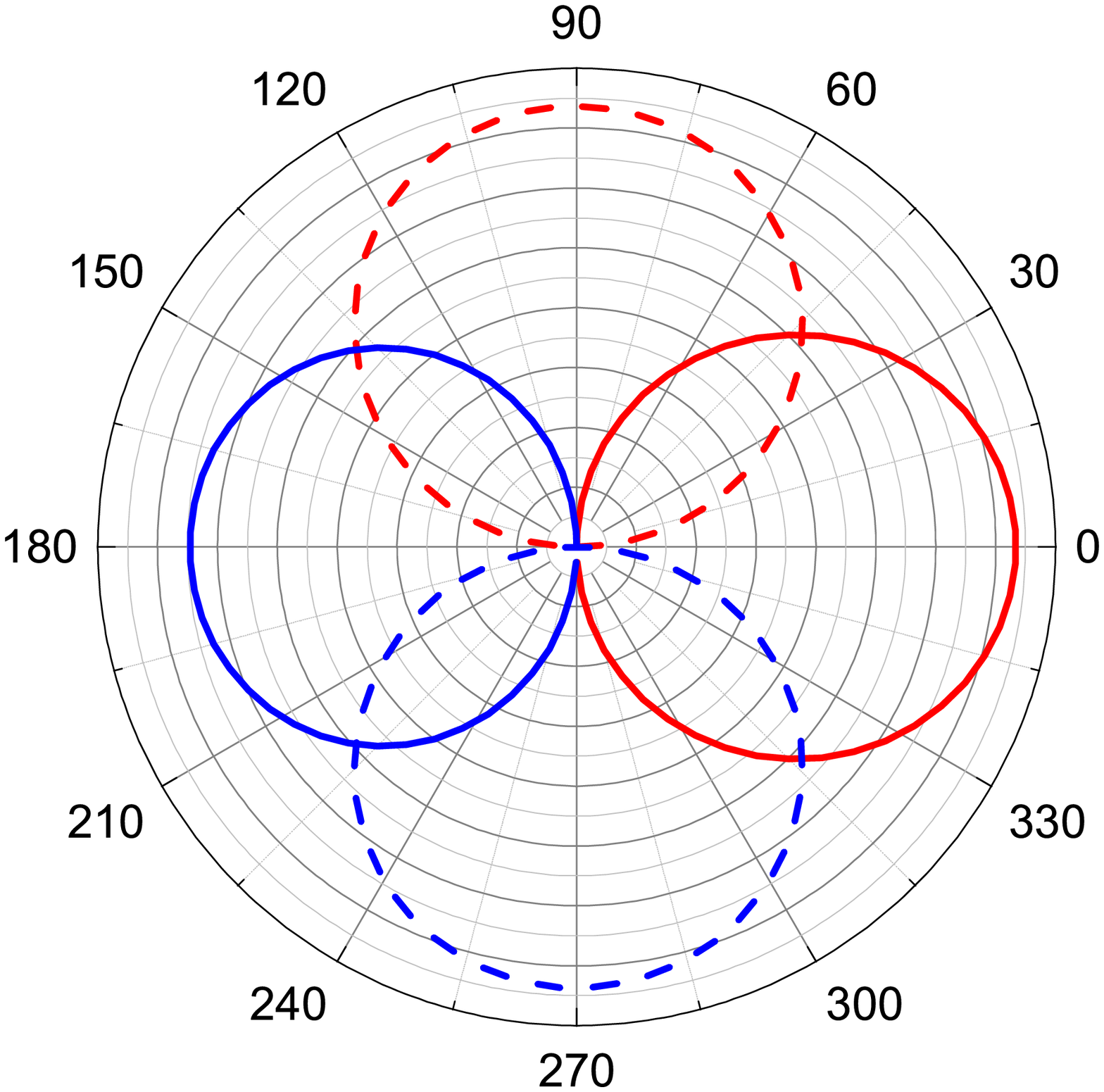}} \\
\vspace{-6 mm}
\raggedright{\small{(1)}}
\end{minipage}
\hfill
\begin{minipage}[h]{0.3\linewidth}
\center{\includegraphics[width=1\linewidth]{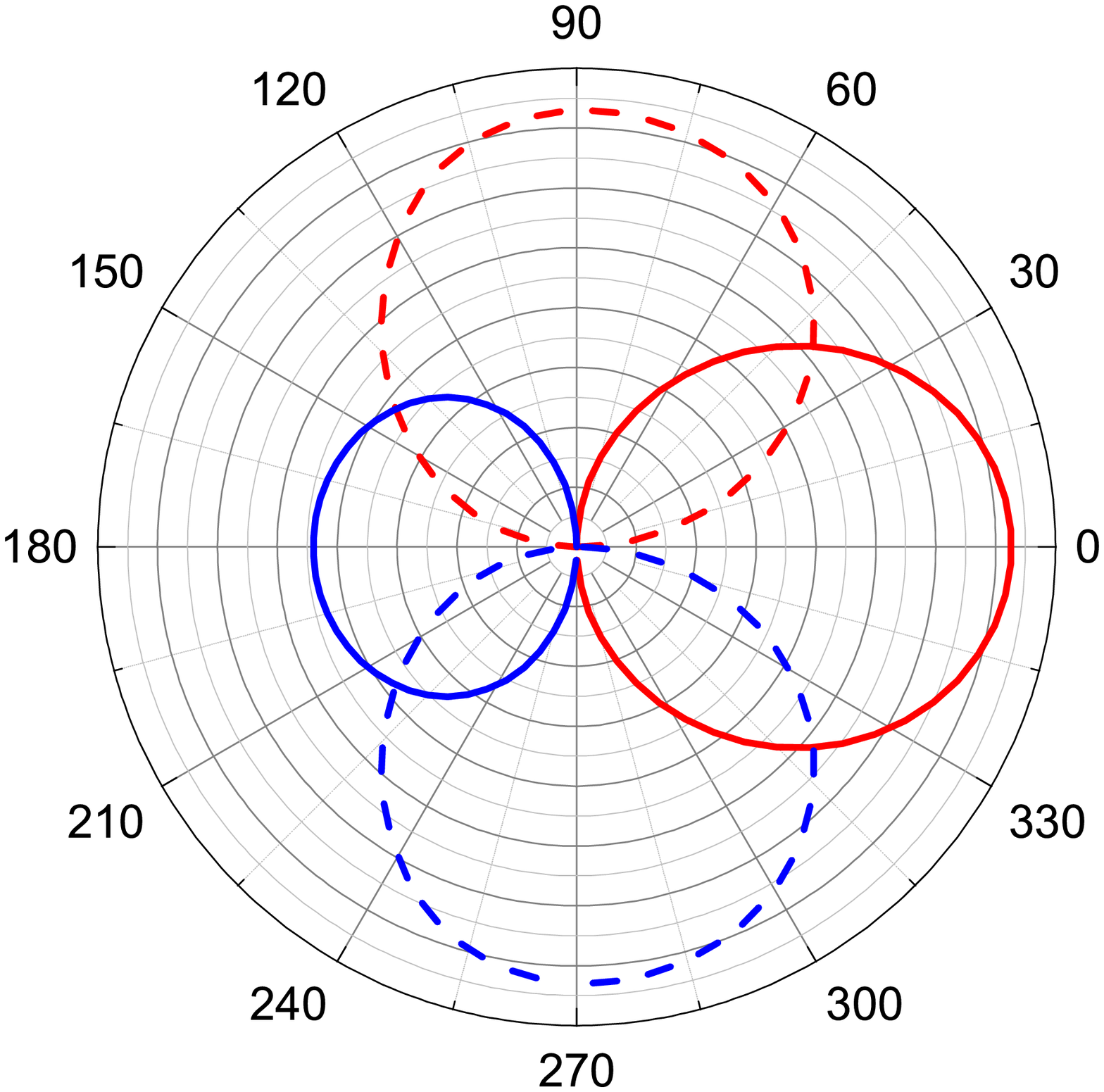}} \\
\vspace{-6 mm}
\raggedright{\small{(2)}}
\end{minipage}
\vfill
\begin{minipage}[h]{0.3\linewidth}
\center{\includegraphics[width=1\linewidth]{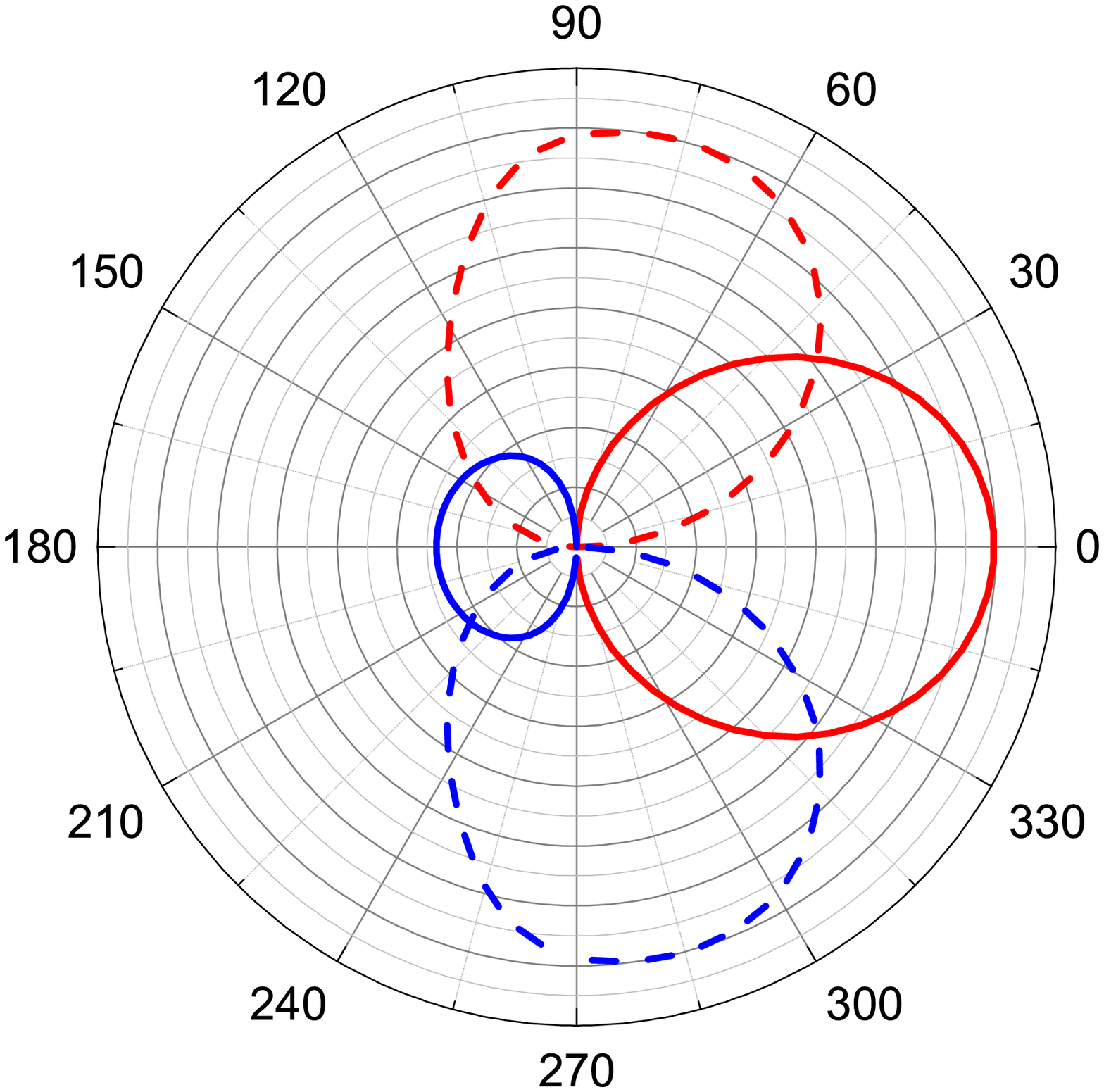}} \\
\vspace{-6 mm}
\raggedright{\small{(3)}}
\end{minipage}
\hfill
\begin{minipage}[h]{0.3\linewidth}
\center{\includegraphics[width=1\linewidth]{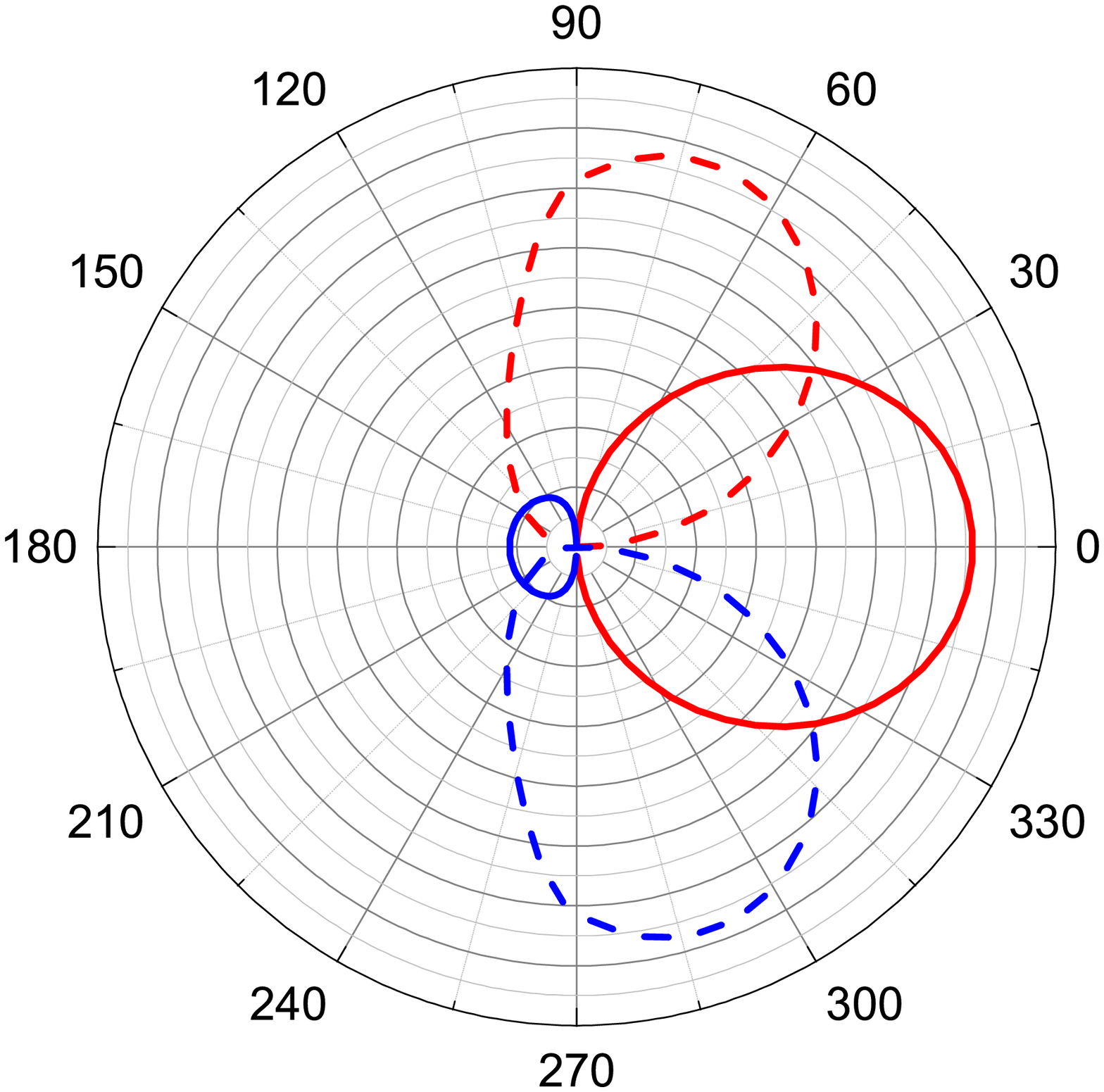}} \\
\vspace{-6 mm}
\raggedright{\small{(4)}}
\end{minipage}
\hfill
\begin{minipage}[h]{0.3\linewidth}
\center{\includegraphics[width=1\linewidth]{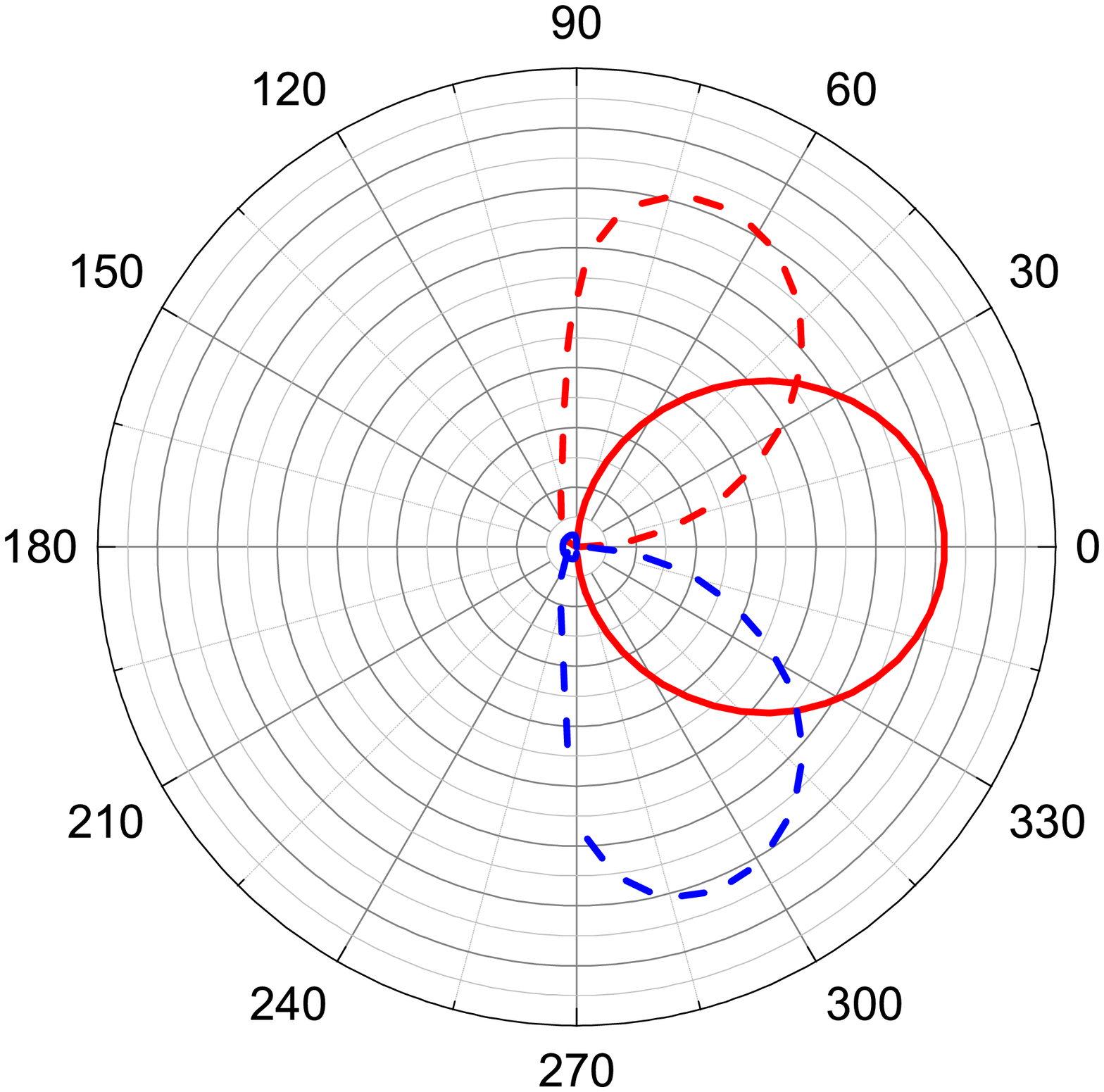}} \\
\vspace{-6 mm}
\raggedright{\small{(4.5)}}
\end{minipage}
\vfill
\begin{minipage}[h]{0.3\linewidth}
\center{\includegraphics[width=1\linewidth]{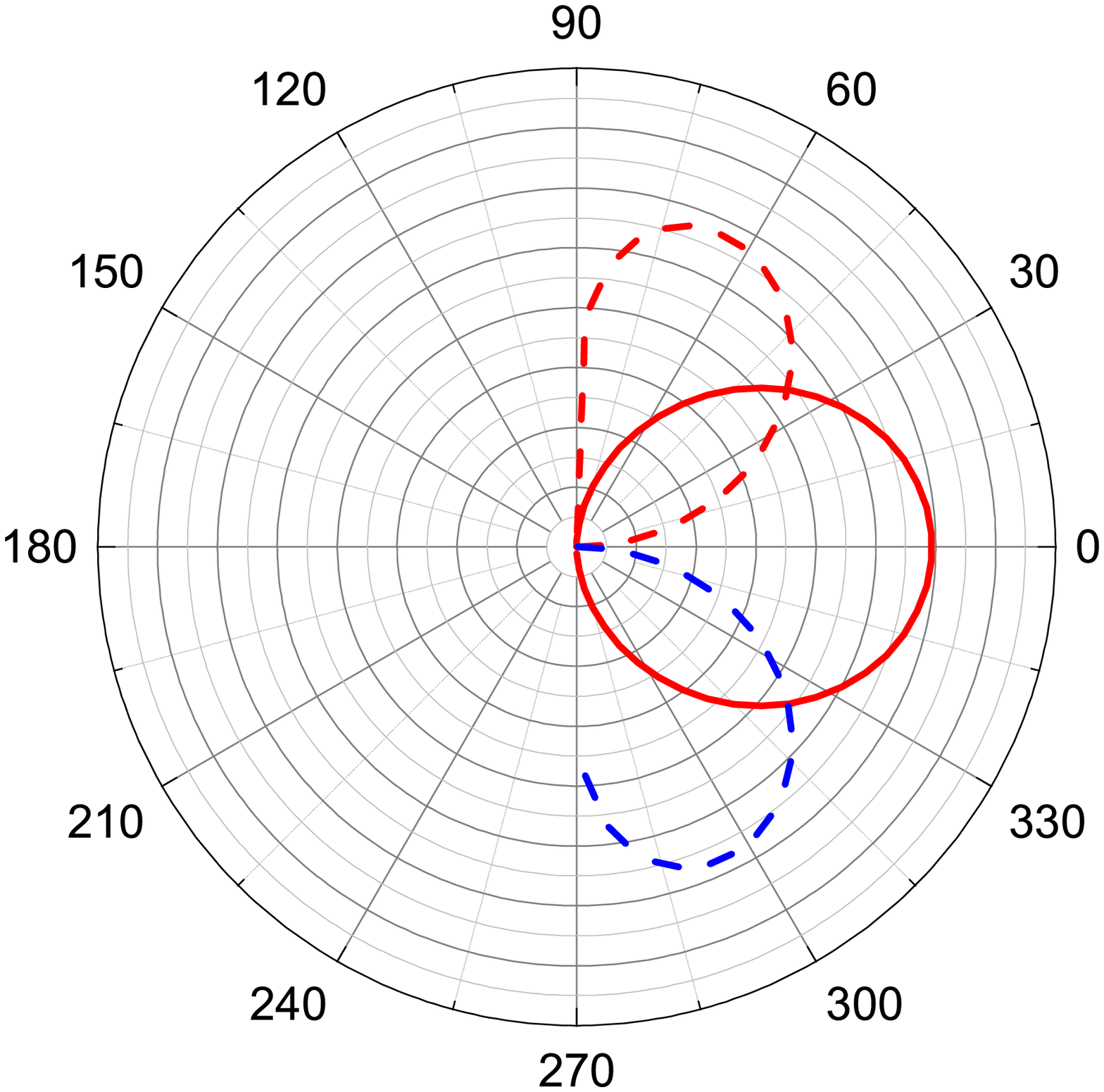}} \\
\vspace{-6 mm}
\raggedright{\small{(5)}}
\end{minipage}
\hfill
\begin{minipage}[h]{0.3\linewidth}
\center{\includegraphics[width=1\linewidth]{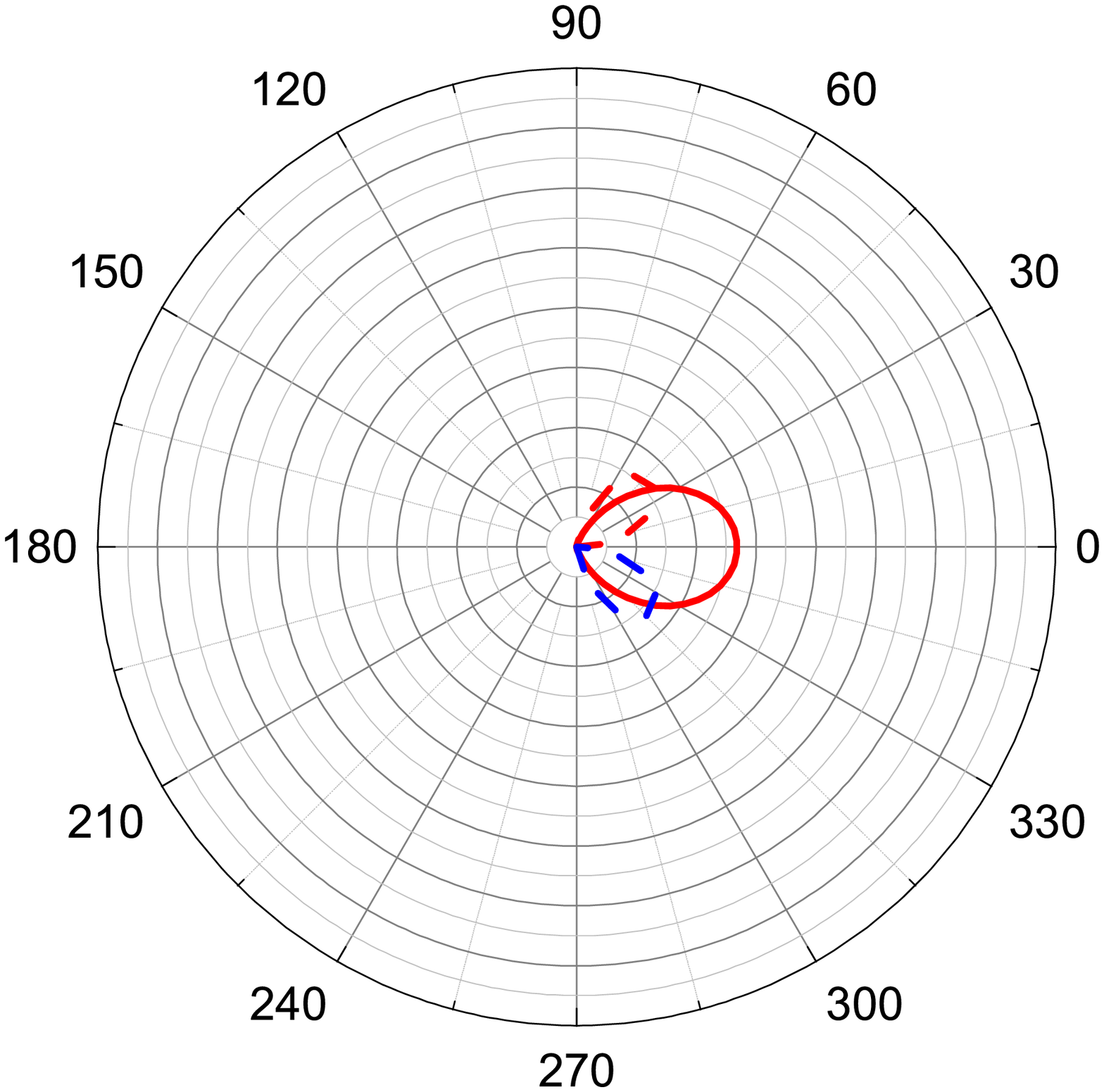}} \\
\vspace{-6 mm}
\raggedright{\small{(6)}}
\end{minipage}
\hfill
\begin{minipage}[h]{0.3\linewidth}
\center{\includegraphics[width=1\linewidth]{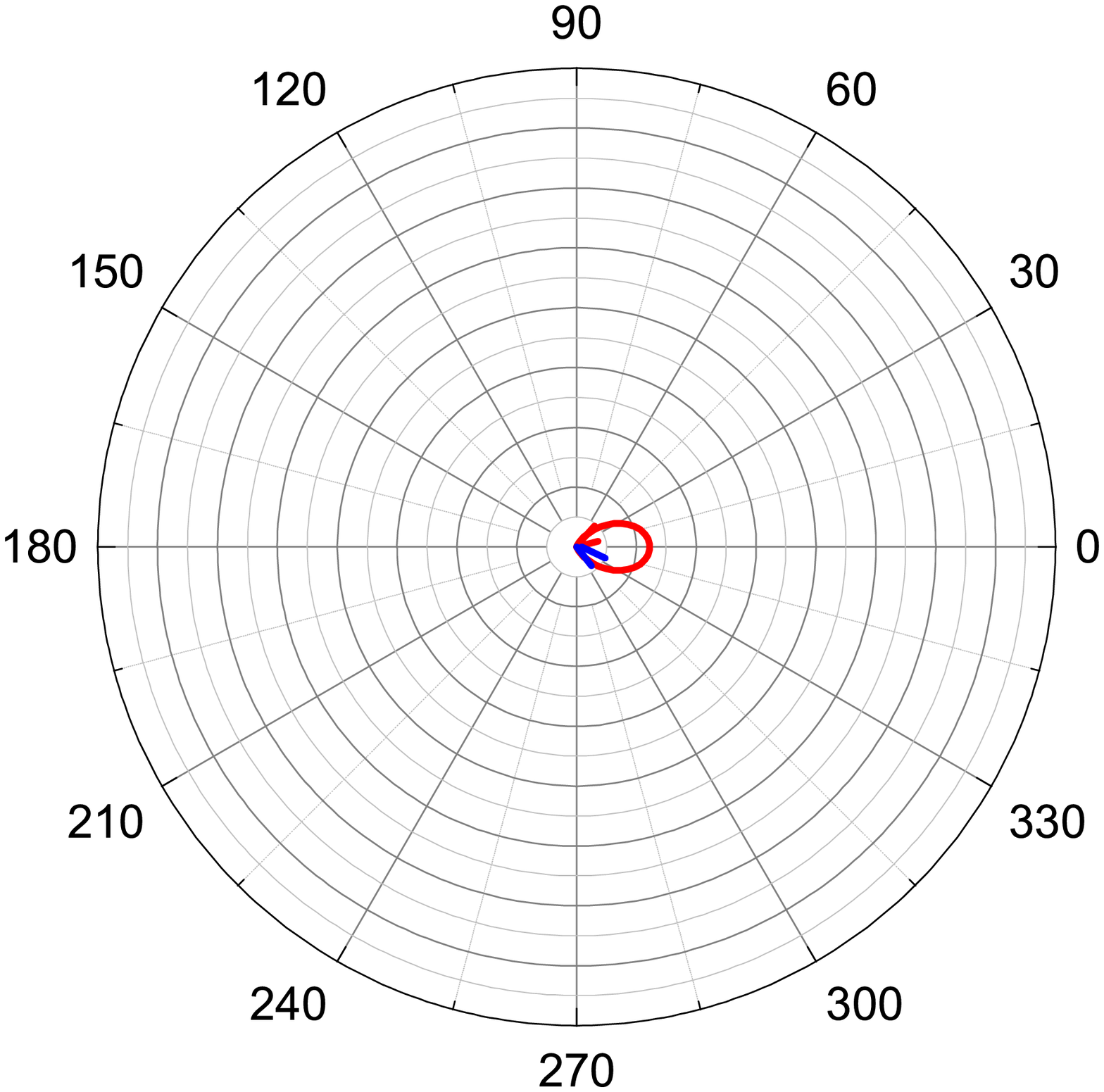}} \\
\vspace{-6 mm}
\raggedright{\small{(7)}}
\end{minipage}

\end{minipage}
\raggedright{d)}
\caption{(Color Online) Spatial distribution of set a) - d) of parameters in
p-wave superconductor with clean metallic surface: \newline
a) the pair potentials $\protect\Delta_x$ and $\protect\Delta_y$ and b) the
surface current density $j_1$, c) contribution of particles with angle $%
\protect\theta$ in the formation of surface current $j_1$ at the different
points of the structure, d) angle dependent pair amplitude $f_1(\protect%
\theta)$ at the different points (0)-(7) of the structure. Set of the points
is following: (0) $x = -2 \protect\xi_0 $, (1) $-0.8 \protect\xi_0 $, (2) $%
-0.4 \protect\xi_0 $, (3) $-0.2 \protect\xi_0 $, (4) $-0.1 \protect\xi_0 $,
(4.5) $-0.03 \protect\xi_0 $, (5) $0 \protect\xi_0 $, (6) $0.25 \protect\xi%
_0 $, (7) $0.5 \protect\xi_0 $. All calculations were performed at $T=0.5
T_C $, $d=\protect\xi_0$ and $l_e=\infty$ .}
\label{f_clean}
\end{figure*}

  At the rough surface, the redistribution of current density takes place: it is pushed towards the border between the clean and the dirty regions (see Fig. \ref{f_rough}). It is also seen from Fig. \ref{f_rough}a that $\Delta_x$ component at this interface is larger than that at specular surface (see Fig. \ref{F_Specular}a), while  $\Delta_y$ is smaller. Due to these competing processes the peak value of the $j_1$ component to the current at $x=0$ is only slightly reduced compared to $j_1 (0)$ calculated in the previous case.

\begin{figure*}[h]
\begin{minipage}[h]{0.5\linewidth}
\begin{minipage}[h]{0.99\linewidth}
\center{\includegraphics[width=1\linewidth]{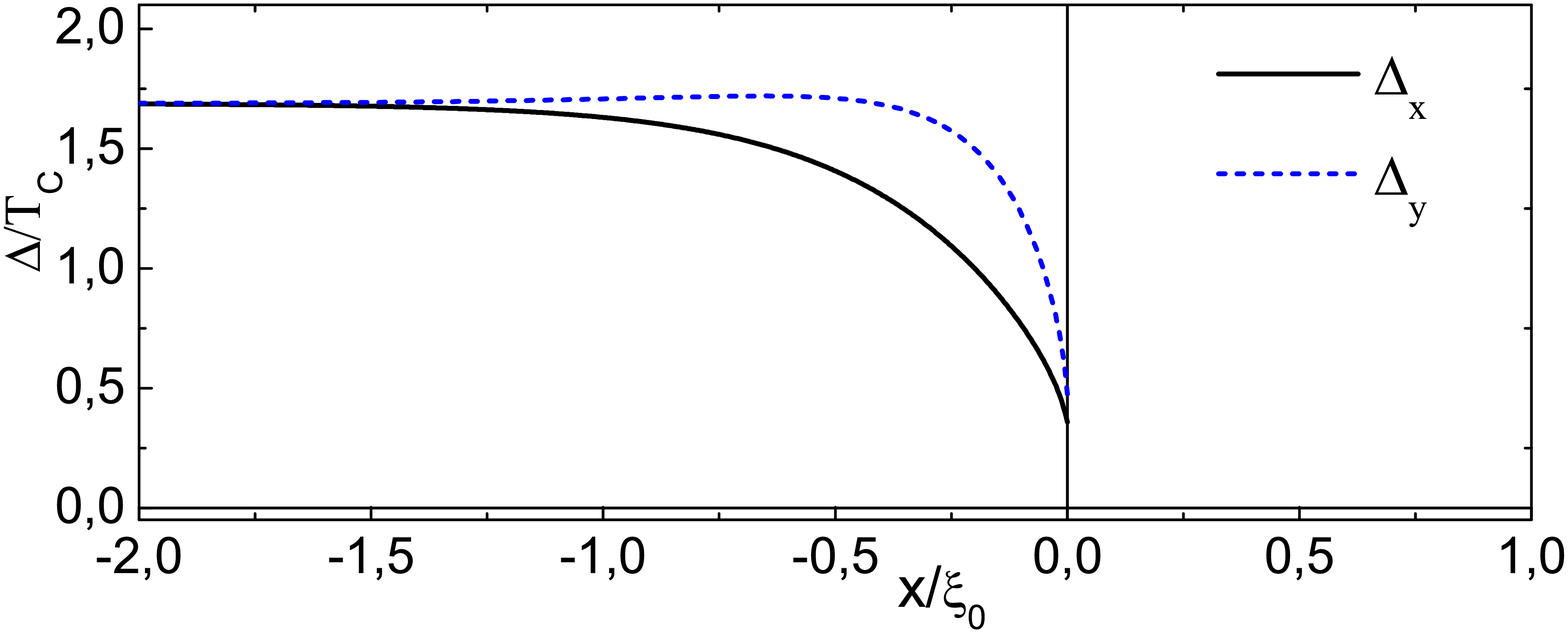}} \\
\vspace{-2 mm}
\raggedright{a)}
\end{minipage}
\vfill
\begin{minipage}[h]{0.99\linewidth}
\center{\includegraphics[width=1\linewidth]{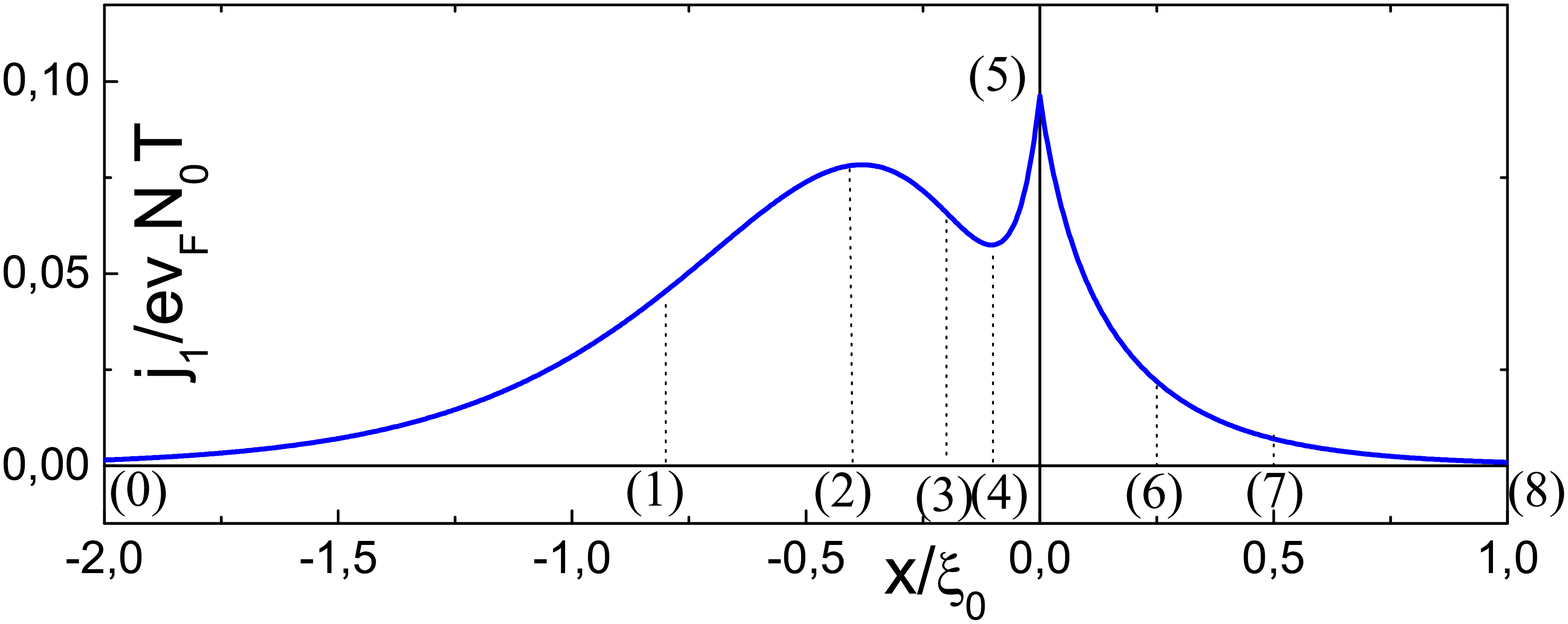}} \\
\vspace{-2 mm}
\raggedright{b)}
\end{minipage}
\end{minipage}
\hfill
\begin{minipage}[h]{0.45\linewidth}
\center{\includegraphics[width=1\linewidth]{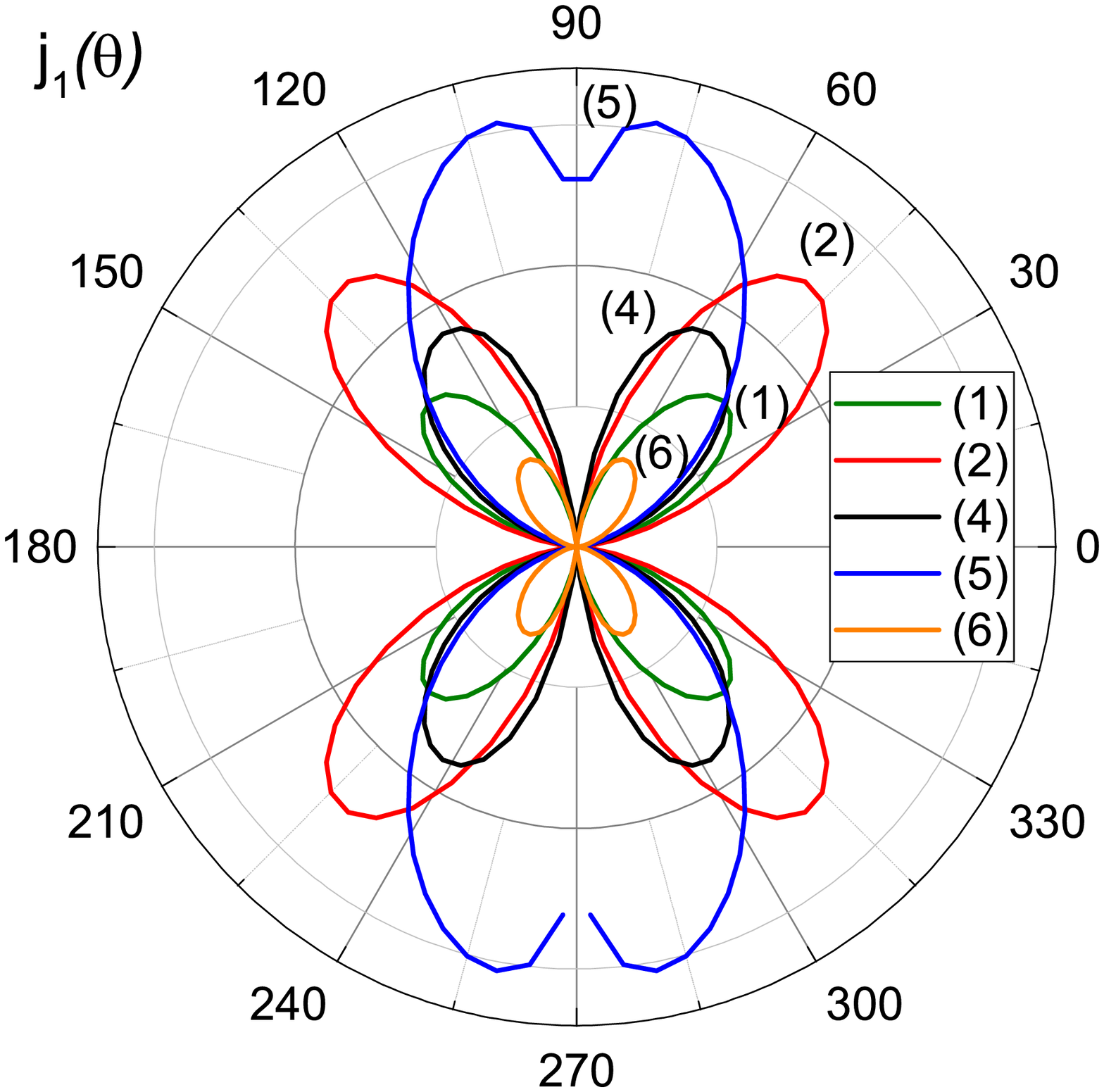}} \\
\vspace{-8 mm}
\raggedright{c)}
\end{minipage}
\vfill
\begin{minipage}[h]{\linewidth}
\begin{minipage}[h]{0.3\linewidth}
\center{\includegraphics[width=1\linewidth]{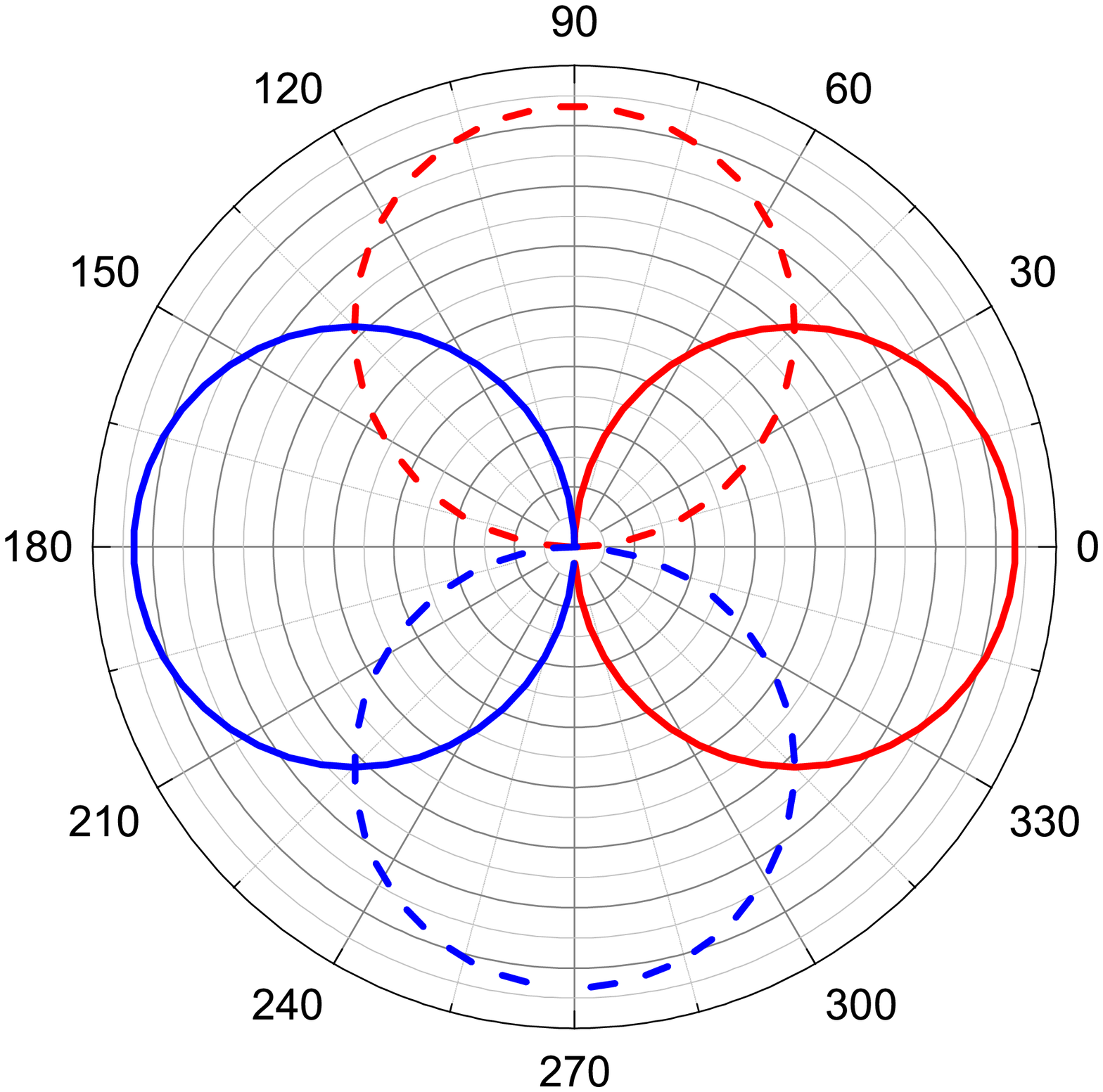}} \\
\vspace{-6 mm}
\raggedright{\small{(0)}}
\end{minipage}
\hfill
\begin{minipage}[h]{0.3\linewidth}
\center{\includegraphics[width=1\linewidth]{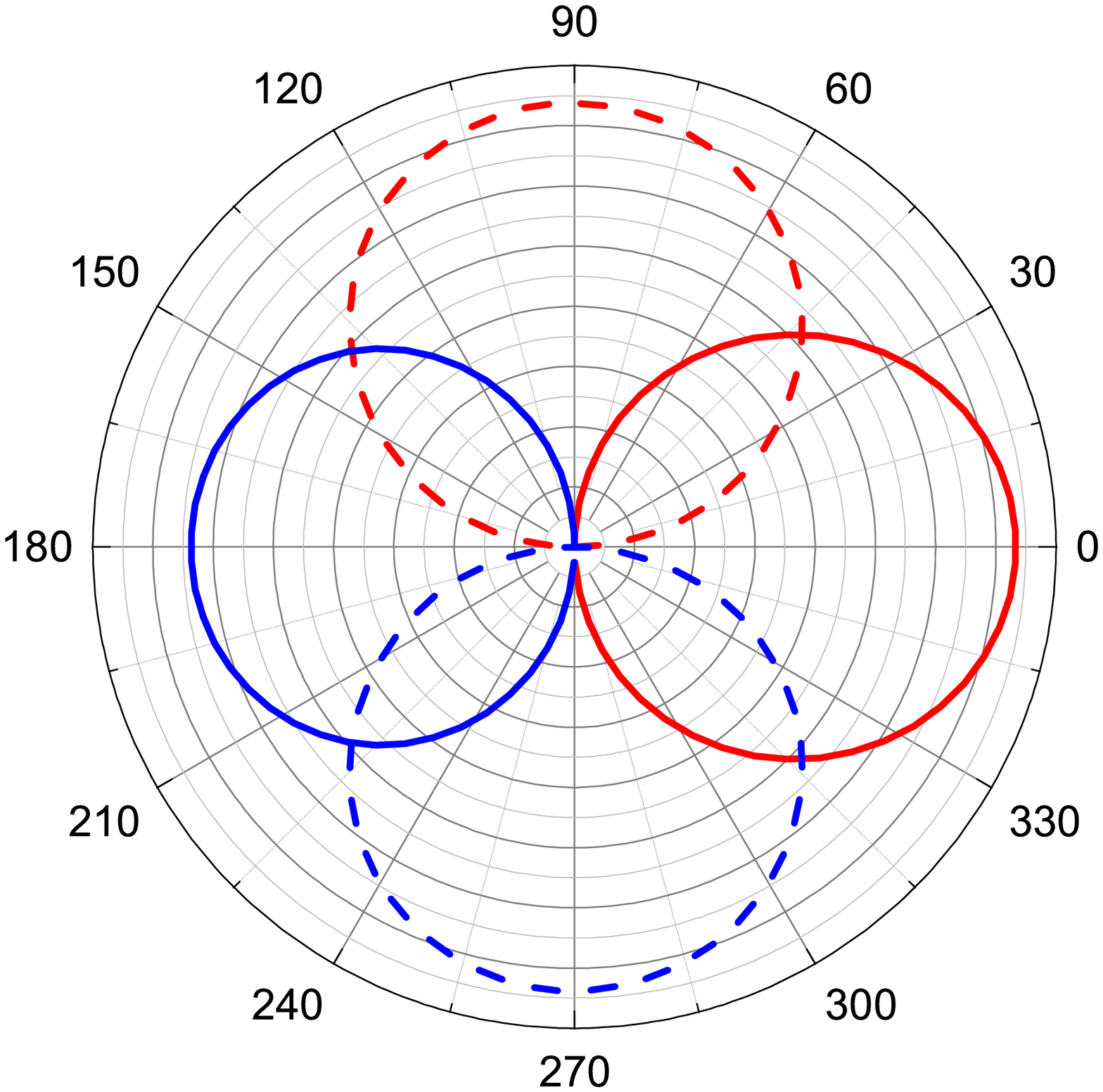}} \\
\vspace{-6 mm}
\raggedright{\small{(1)}}
\end{minipage}
\hfill
\begin{minipage}[h]{0.3\linewidth}
\center{\includegraphics[width=1\linewidth]{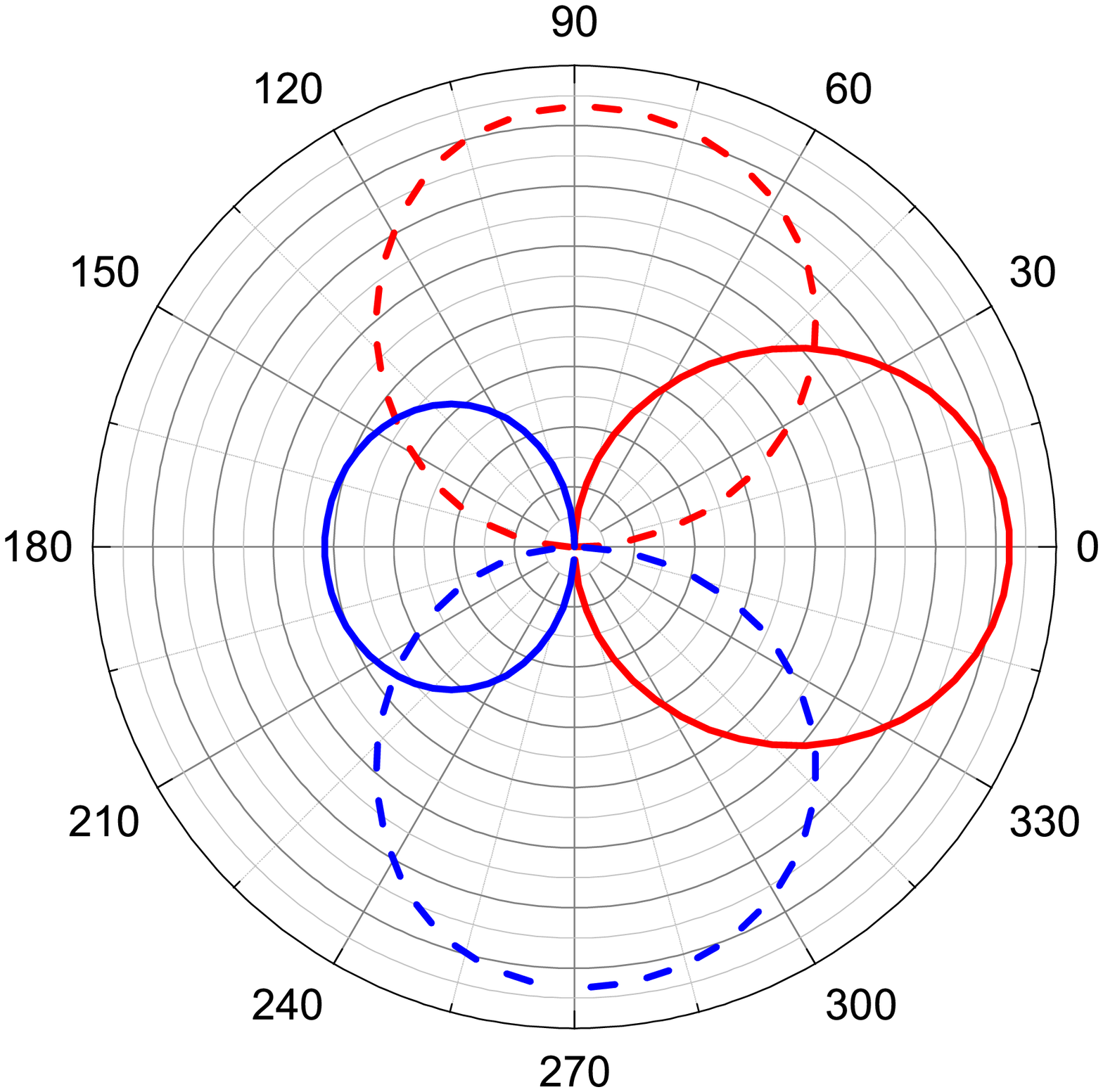}} \\
\vspace{-6 mm}
\raggedright{\small{(2)}}
\end{minipage}
\vfill
\begin{minipage}[h]{0.3\linewidth}
\center{\includegraphics[width=1\linewidth]{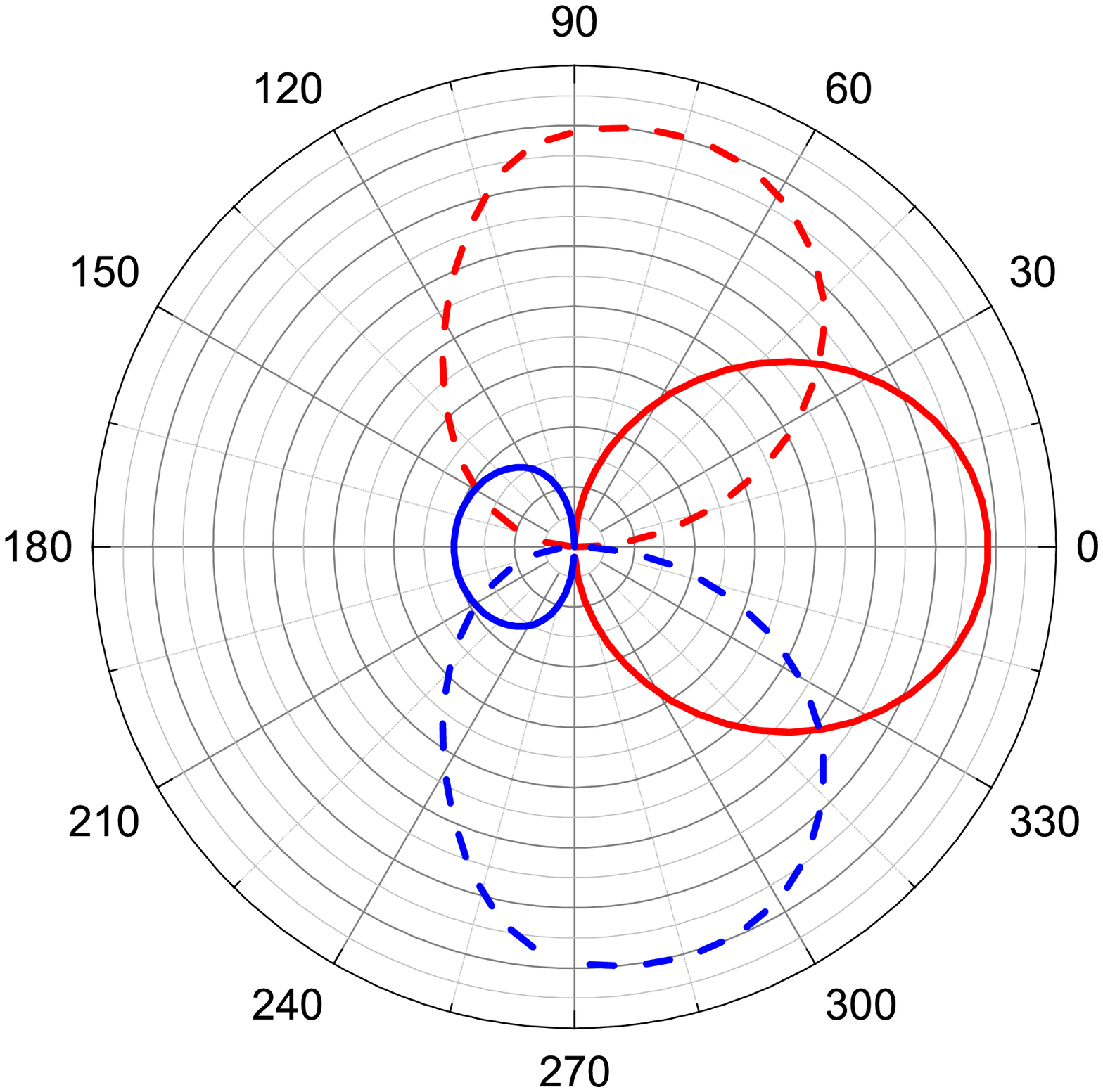}} \\
\vspace{-6 mm}
\raggedright{\small{(3)}}
\end{minipage}
\hfill
\begin{minipage}[h]{0.3\linewidth}
\center{\includegraphics[width=1\linewidth]{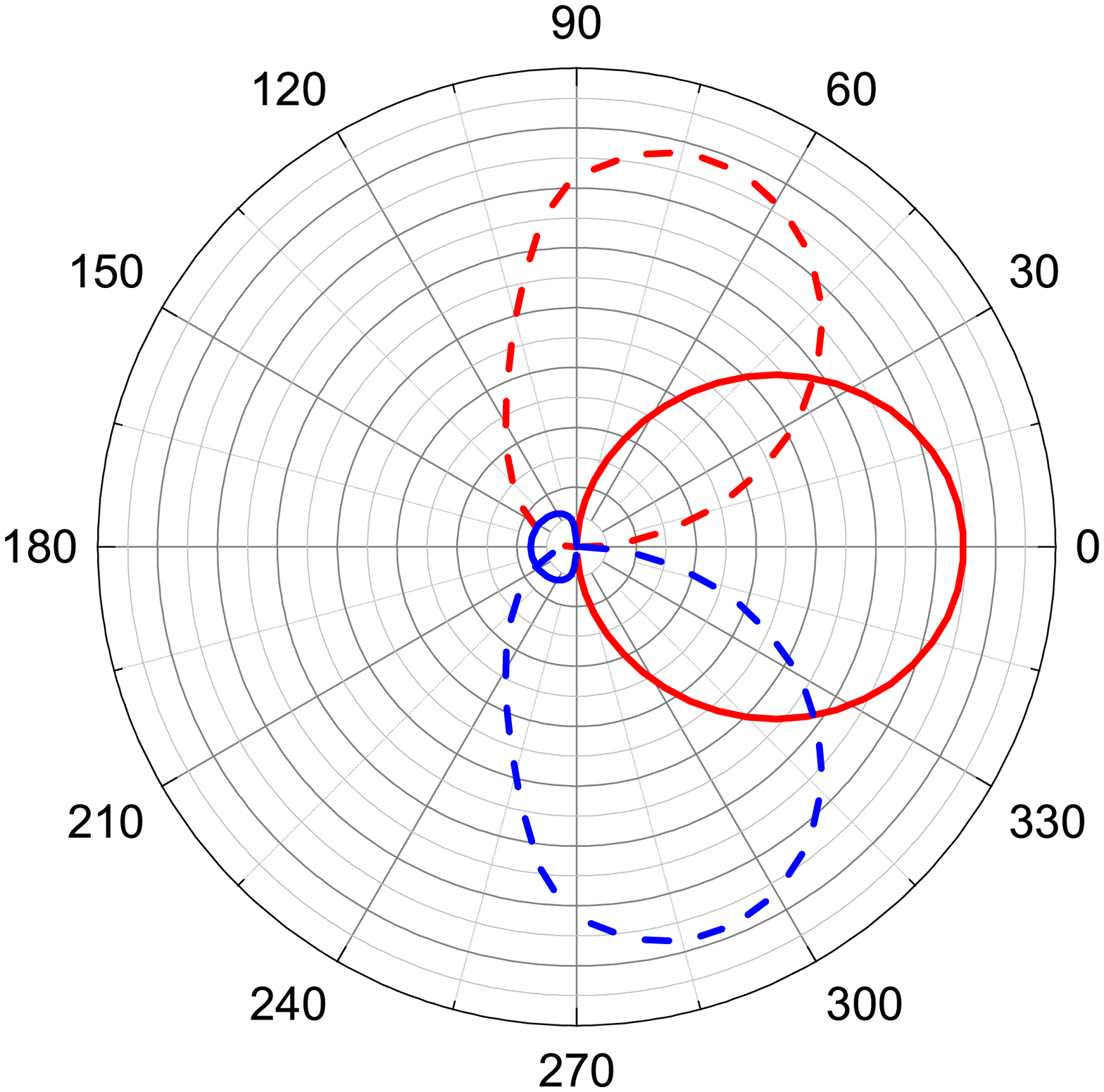}} \\
\vspace{-6 mm}
\raggedright{\small{(4)}}
\end{minipage}
\hfill
\begin{minipage}[h]{0.3\linewidth}
\center{\includegraphics[width=1\linewidth]{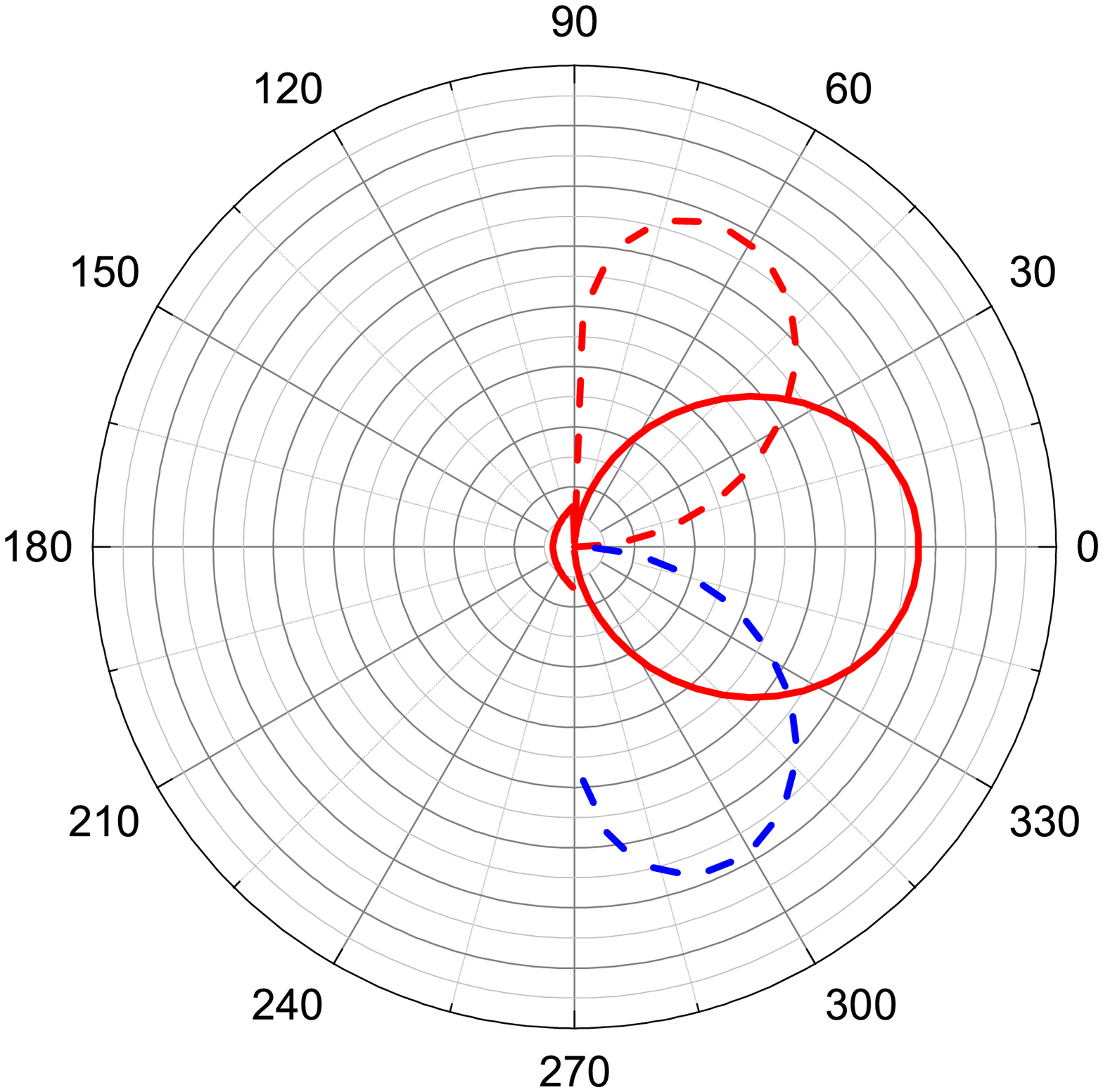}} \\
\vspace{-6 mm}
\raggedright{\small{(5)}}
\end{minipage}
\vfill
\begin{minipage}[h]{0.3\linewidth}
\center{\includegraphics[width=1\linewidth]{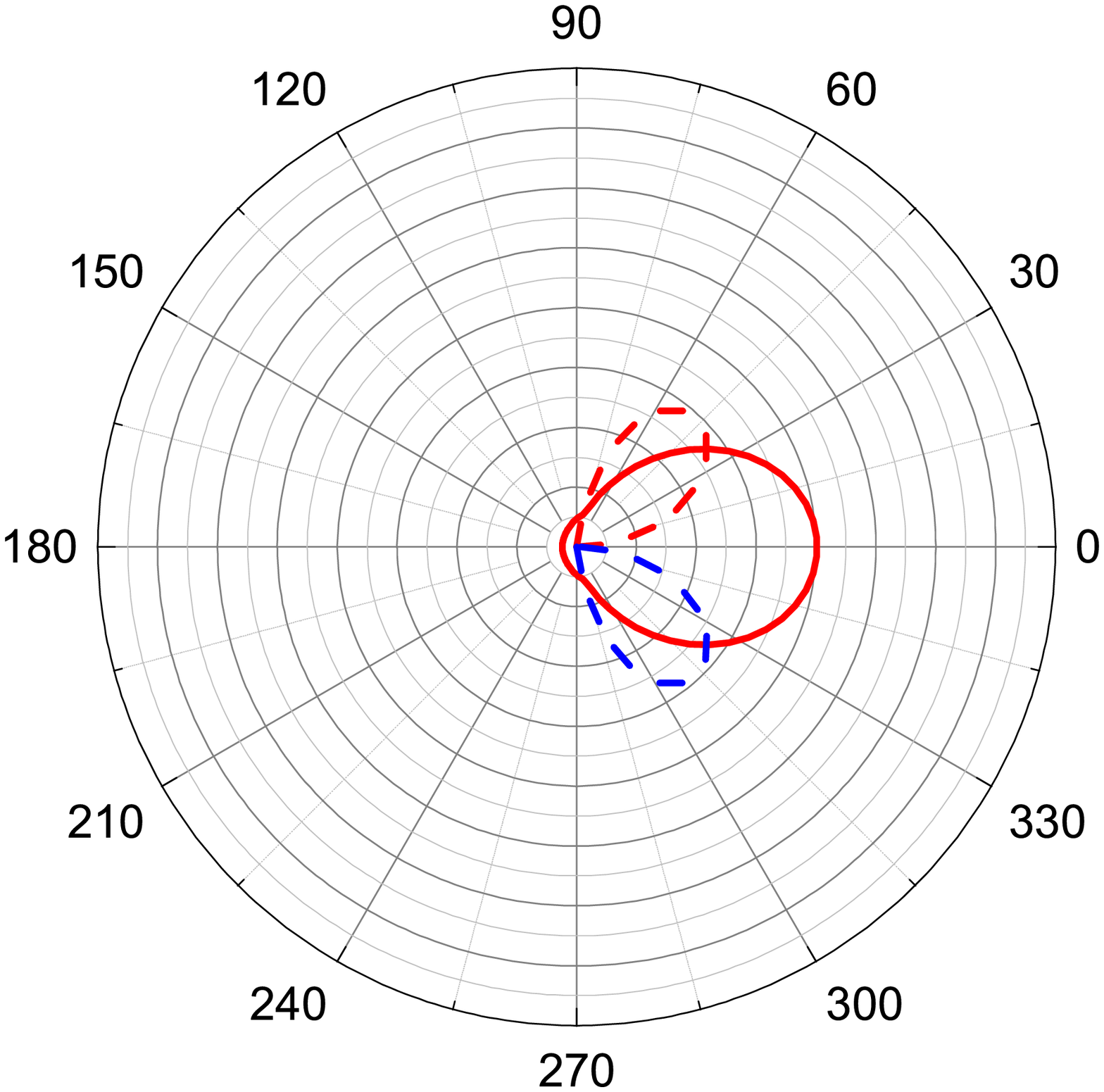}} \\
\vspace{-6 mm}
\raggedright{\small{(6)}}
\end{minipage}
\hfill
\begin{minipage}[h]{0.3\linewidth}
\center{\includegraphics[width=1\linewidth]{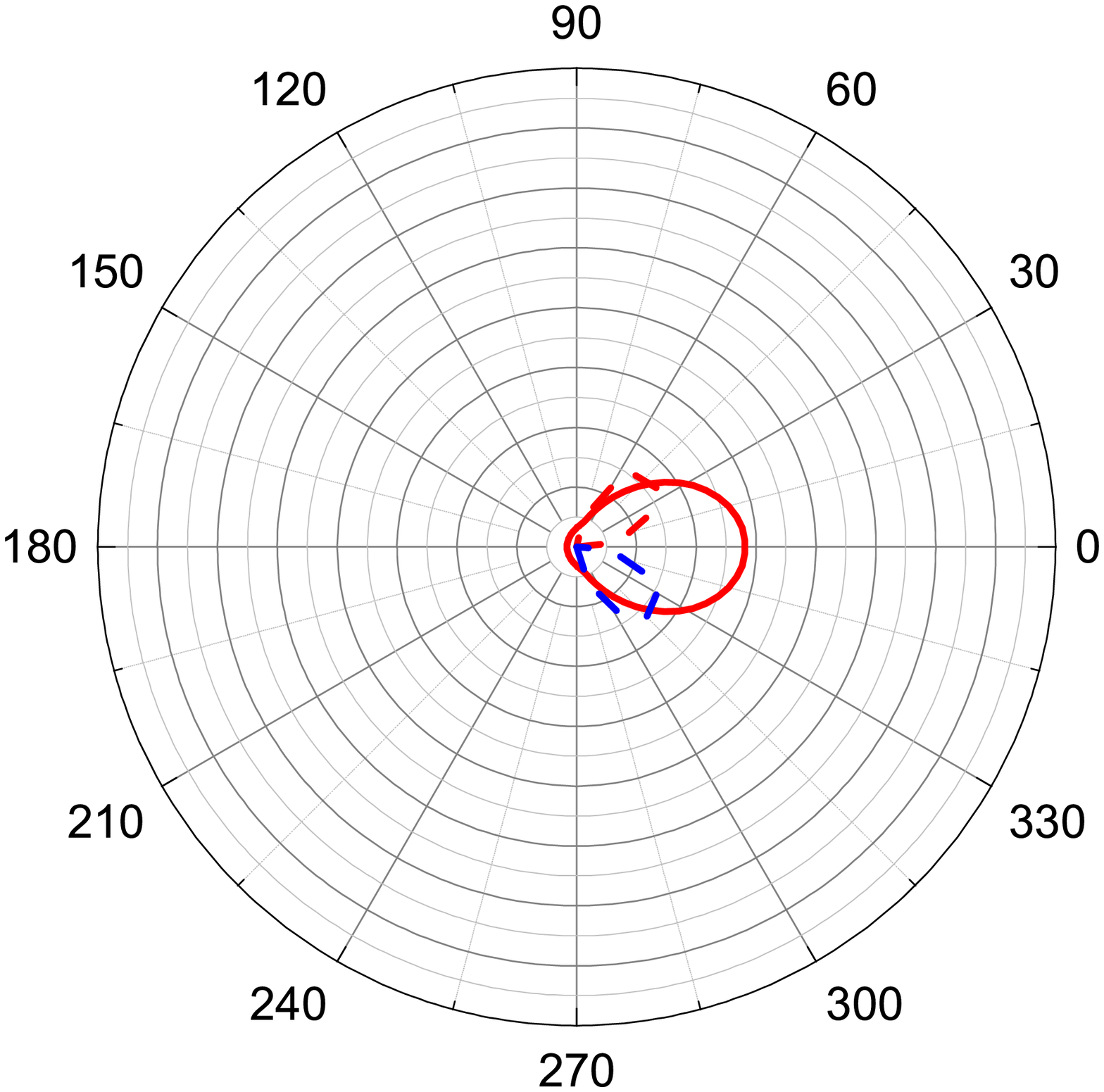}} \\
\vspace{-6 mm}
\raggedright{\small{(7)}}
\end{minipage}
\hfill
\begin{minipage}[h]{0.3\linewidth}
\center{\includegraphics[width=1\linewidth]{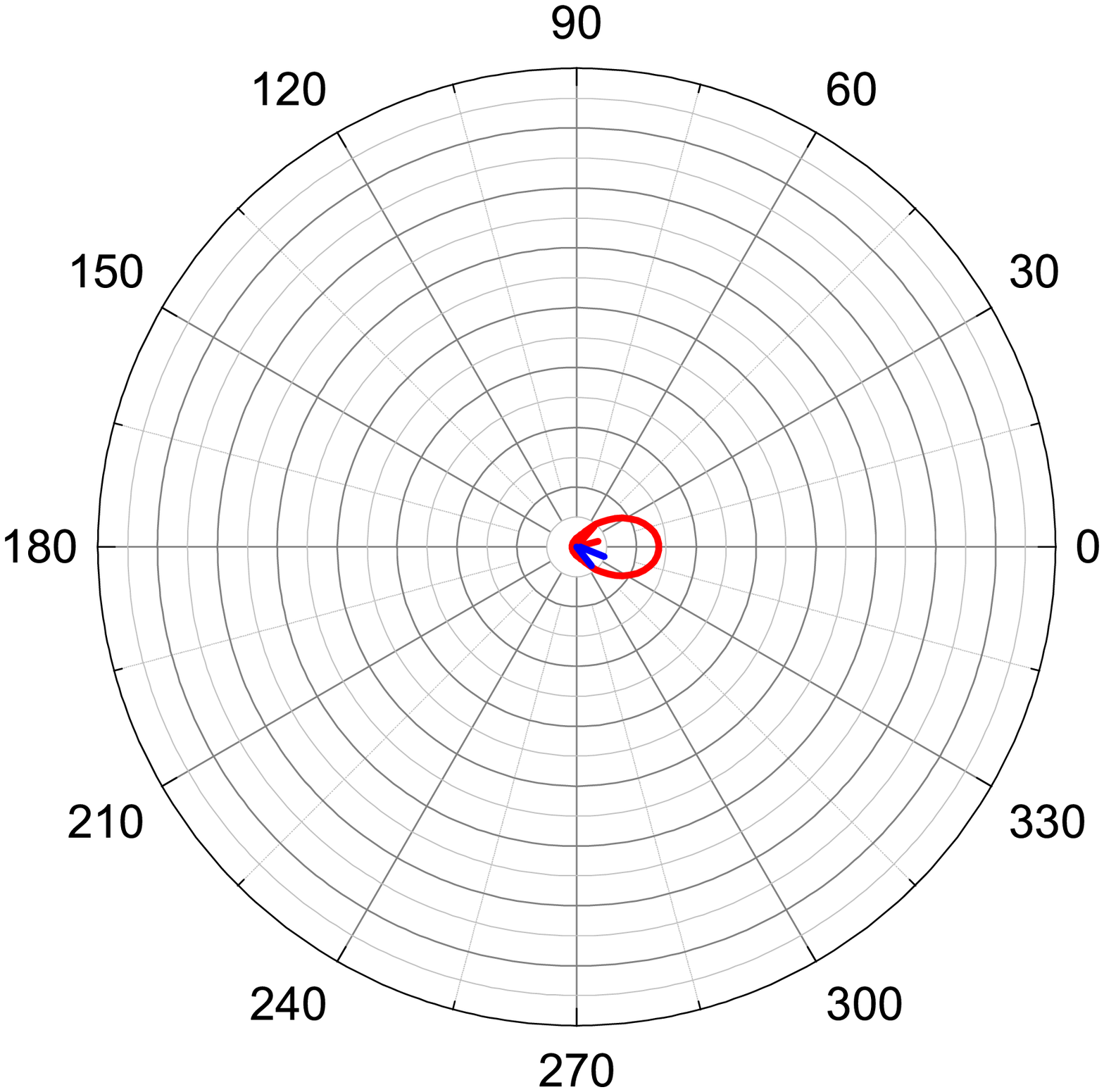}} \\
\vspace{-6 mm}
\raggedright{\small{(8)}}
\end{minipage}

\end{minipage}
\raggedright{d)}
\caption{(Color Online) Spatial distribution of set a) - d) of parameters in
p-wave superconductor with diffusive metallic surface: \newline
a) the pair potentials $\protect\Delta_x$ and $\protect\delta_y$ and b) the
surface current density $j_1$, c) contribution of particles with angle $%
\protect\theta$ in the formation of surface current $j_1$ at the different
points of the structure, d) angle dependent pair amplitude $f_1(\protect%
\theta)$ at the different points (0)-(8) of the structure. Set of the points
is following: (0) $x = -2 \protect\xi_0 $, (1) $-0.8 \protect\xi_0 $, (2) $%
-0.4 \protect\xi_0 $, (3) $-0.2 \protect\xi_0 $, (4) $-0.1 \protect\xi_0 $,
(5) $0 \protect\xi_0 $, (6) $0.25 \protect\xi_0 $, (7) $0.5 \protect\xi_0 $,
(8) $x = 1 \protect\xi_0 $. All calculations were performed at $T=0.5 T_C$, $%
d= l_e$ and $\protect\xi_0=0.3 l_e$.}
\label{f_inter}
\end{figure*}

The situation changes drastically when the thickness of the normal layer on the top of p-wave superconductor exceeds $\xi_0$ irrespective of the degree of purity of the normal metal. As follows from Fig. \ref{F_Specular}b and Fig. \ref{f_inter}b, in both considered cases there is nearly one order of magnitude suppression of maximum value of $j_1$ compare to $j_1 (0)$ at specular interface. This result is correlated with that obtained in \cite{Lederer} in the tight-binding model for the clean N metal. It is also seen from Fig. \ref{F_Specular}b and Fig. \ref{f_inter}b that spatial distribution of the density of spontaneous current for the clean and dirty cases are completely different. 

The difference in the current distributions considered above can be accessed experimentally by muon spin rotation technique \cite{Flokstra}. According to the results of our study, such measurements of surface currents may provide important information on the underlying pairing symmetry in the bulk.

Acknowledgements. The authors acknowledge fruitful discussions with Y. Asano and Y. Tanaka. This work was partially supported by RFBR-JSPS grants 15-52-50054, 17-52-50080, RFBR grant 16-29-09515-ofi-m, by Russian Science Foundation, Project No. 15-12-30030, and by Ministry of Education and Science of the Russian Federation, grants MK-5813.2016.2 and 14Y26.31.0007.

\bigskip

\bigskip

\bigskip


\end{document}